\documentstyle[eqsecnum,preprint,epsfig,aps]{revtex}
\tighten
\draft
\widetext
\input epsf
\bigskip
\bigskip
\begin{document}
%\date{\today}
\bigskip
\medskip
\title{A Cosmology of the Brane World}
\author{\'E.\ \'E.\ Flanagan, S.-H. H. Tye, and I. Wasserman}
\address{Laboratory for Nuclear Studies and Center for Radiophysics
and Space Research\\ Cornell University, Ithaca, NY 14853.}
\date{\today}
\maketitle
\def\NPB#1#2#3{Nucl. Phys. B{#1} (19#2) #3}
\def\vt{\vartheta}
\def\preal{{\rm Re\,}}
\def\pim{{\rm Im\,}}
\def\ds{\displaystyle}
\def\yzero{\smash{\hbox{$y\kern-4pt\raise1pt\hbox{${}^\circ$}$}}}
\def\p{\partial}
\def\a{\alpha}
\def\b{\beta}
\def\g{\gamma}
\def\d{\delta}
\def\Om{\Omega}
\def\om{\omega}
\def\th{\theta}
\def\psivec{{\mbox{\boldmath $\psi$}}}
\def\vt{\vartheta}
\def\vphi{\varphi}
\def\-{\hphantom{-}}
\def\ov{\overline}
\def\s2{\frac{1}{\sqrt2}}
\def\wh{\widehat}
\def\wt{\widetilde}
\def\oh{\frac{1}{2}}
\def\beq{\begin{equation}}
\def\eeq{\end{equation}}
\def\beqa{\begin{eqnarray}}
\def\eeqa{\end{eqnarray}}
\def\tr{{\rm tr \,}}
\def\Tr{{\rm Tr \,}}
\def\diag{{\rm diag \,}}
\def\IF{\relax{\rm I\kern-.18em F}}
\def\II{\relax{\rm I\kern-.18em I}}
\def\IP{\relax{\rm I\kern-.18em P}}
\def\IC{\relax\hbox{\kern.25em$\inbar\kern-.3em{\rm C}$}}
\def\IR{\relax{\rm I\kern-.18em R}}
\def\hm{\relax{n_H}}
\def\vac{|0 \rangle}
\def\vm{\relax{n_V}}
\def\cc{{\cal C}}
\def\ck{{\cal K}}
\def\ci{{\cal I}}
\def\cu{{\cal U}}
\def\cg{{\cal G}}
\def\cn{{\cal N}}
\def\cam{{\cal M}}
\def\mplI{M_{\rm Pl, I}}
\def\cp{{\cal P}}
\def\ct{{\cal T}}
\def\cv{{\cal V}}
\def\cz{{\cal Z}}
\def\ch{{\cal H}}
\def\cf{{\cal F}}
\def\tv{\tilde v}
\def\Dsl{\,\raise.15ex\hbox{/}\mkern-13.5mu D} %this one can be subscripted
\def\rzero{r_0}
\def\fI{f_{\rm I}}
\def\IZ{Z\kern-.4em  Z}
\def\id{{\rm I}}
\def\bP{{\bf\rm P}}
\def\bF{{\bf\rm F}}
\def\bZ{{\bf\rm Z}}
\def\l{{\lambda}}
\def\ep{{\epsilon}}
\def\mI{m_{\rm I}}
\def\rI{r_{\rm I}}
\def\lta{\mathbin{\lower 3pt\hbox
  {$\rlap{\raise 5pt\hbox{$\char'074$}}\mathchar"7218$}}} %< or of order
\def\gta{\mathbin{\lower 3pt\hbox
	{$\rlap{\raise 5pt\hbox{$\char'076$}}\mathchar"7218$}}} %> or of order

\def\lta{\mathbin{\lower 3pt\hbox
  {$\rlap{\raise 5pt\hbox{$\char'074$}}\mathchar"7218$}}} %< or of order
\def\gta{\mathbin{\lower 3pt\hbox
  {$\rlap{\raise 5pt\hbox{$\char'076$}}\mathchar"7218$}}} %> or of order
\def\psivec{{\mbox{\boldmath $\psi$}}}
\def\grad{{\mbox{\boldmath $\nabla$}}}
\def\nablab{\hat\nabla}
\def\gradb{\hat{\grad}}
\def\cdotvec{{\mbox{\boldmath $\cdot$}}}
\def\rzero{r_0}
\def\taub{\hat{\tau}}
\def\tb{\hat{t}}
\def\ab{\hat{a}}
\def\Hb{\hat{H}}
\def\sb{\hat{s}}
\def\xvec{{\bf x}}
\def\mpl{M_{\rm Pl}}
\def\be{\begin{equation}}
\def\ee{\end{equation}}
\def\baray{\begin{eqnarray}}
\def\earay{\end{eqnarray}}
\def\Vhatz{\hat V_0}
\def\Vrzero{\hat V_{r,0}}
\def\Vhat1{\hat V_1}
\def\mplI{M_{\rm Pl, I}}
\def\mI{m_{\rm I}}
\def\rI{r_{\rm I}}
\def\Hbar{\hat{H}}
\def\Tc1{T_{{\rm c},1}}
\def\vbulk{V_{\rm bulk}(r)}
\def\Ubulk{U_{\rm bulk}}
\def\ve{V_E}
\def\Upsi{\Upsilon_i}
\def\fI{f_{\rm I}}
\def\veff{V_{\rm eff}}
\def\bh{\hat{h}}
\def\xb{\hat{x}}
\def\xbvec{{\bf{\xb}}}
\def\Uvec{{\bf U}}
\def\phih{\phi_H}
\def\phib{\hat{\phi}}
\def\phihb{\phib_H}
\def\phihbp{\phihb^\prime}
\def\phihbdp{\phihb^{\prime\prime}}
\def\etap{\eta^\prime}
\def\etadp{\eta^{\prime\prime}}
\def\bhp{\bh^\prime}
\def\bhdp{\bh^{\prime\prime}}
\def\rhobar{\hat{\rho}}
\def\psitp{\psi^{\prime\prime\prime}}
\def\psidp{\psi^{\prime\prime}}
\def\psip{\psi^\prime}
\def\Gev{{\rm GeV}}
\def\Tev{{\rm TeV}}
\def\rco{\rho_{c,0}}
\def\arh{a_{rh}}
\def\trh{T_{rh}}
\def\rorh{\rh_{rh}}
\def\scal{{\cal S}}
\def\so{\scal_0^{1/3}}
\def\srh{\scal_{rh}^{1/3}}
\def\aexp{a_{exp}}
\def\epsrh{\epsilon_{rh}}
\def\ecal{{\cal E}}
\def\fcal{{\cal F}}
\def\erh{\ecal_{rh}}
\def\ahub{a_{Hubble}}
\def\arhone{a_{rh,1}}
\def\arhtwo{a_{rh,2}}
\def\aexpone{a_{exp,1}}
\def\aexptwo{a_{exp,2}}
\def\srhone{\scal_{rh,1}^{1/3}}
\def\srhtwo{\scal_{rh,2}^{1/3}}
\def\erhone{\ecal_{rh,1}}
\def\erhtwo{\ecal_{rh,2}}
\def\epsrhone{\epsilon_{rh,1}}
\def\epsrhtwo{\epsilon_{rh,2}}
\def\trhone{T_{rh,1}}
\def\trhtwo{T_{rh,2}}

\begin{abstract} We develop a possible cosmology for a Universe in which
there are $n$ additional spatial dimensions of variable scale, and an
associated scalar field, the radion, which is distinct from the field
responsible for inflation, the inflaton. Based on
a particular {\it ansatz} for the effective potential for the inflaton
and radion (which may emerge in string
theory), we show that the early expansion of the Universe may proceed
in three stages. During the earliest phase, the radion field becomes
trapped at a value much smaller than the size of the extra dimensions
today. Following this phase, the Universe expands exponentially, but
with a Planck mass smaller than its present value. Because 
the Planck mass during inflation is small, we find
that density fluctuations in agreement with observations can
arise naturally.
When inflation 
ends, the Universe reheats, and the radion becomes free to expand
once more. During the third phase the Universe is
``radiation-dominated'' and tends toward a fixed-point evolutionary
model in which  
the radius of the extra dimension grows, but the temperature remains
unchanged. Ultimately, the radius of the extra dimensions becomes
trapped once again at its present value, and a short period of
exponential expansion, which we identify with the electroweak
phase transition, ensues. Once this epoch is over, the Universe
reheats to a temperature $\lta m_{EW}$, the electroweak scale,
and the mature Universe evolves according to standard cosmological
models. We show that the present day energy density in radions
can be smaller than the closure density of the Universe if the
second inflationary epoch lasts $\sim 8$ e-foldings or more; the
present-day radion mass turns out to be small ($m_{\rm radion}\lta$ eV,
depending on parameters). We argue that although our model 
envisages considerable time evolution in the Planck mass, substantial
spatial fluctuations in Newton's constant are not produced.
\end{abstract}

\vfill\eject

\section{Introduction}

Recently it was suggested that the fundamental scale of gravity
may be as low as TeV \cite{add}. According to this idea,
the observed weakness of gravity 
is associated with $n$ new, relatively large spatial dimensions 
(compactified to a size $\sim r_0$) in which only gravity can propagate.
In this picture, all the standard model particles live in a set of branes
with three extended space dimensions (``brane modes''), while gravitons 
live in the higher dimensional bulk of spacetime (``bulk modes''). 
This scenario turns out to be quite natural 
in (Type I) string theory. In this "brane world" picture \cite{aadd,bw}, 
the standard model particles are open strings whose ends must end on 
the branes ({\em e.g.}, stretched between branes), 
while gravitons are closed string states that can move away from the 
branes and into the bulk.
The relation between today's Planck scale $\mpl=1.2$ x $10^{19}$ 
GeV and the fundamental string scale $m_s$ is approximately given by
\begin{equation}
\mpl^2 \sim m_s^{n + 2}r_0^n
\label{planckscale}
\end{equation}
Phenomenological and astrophysical constraints imply that $m_s$ may be 
as low as a few TeV, with $n \ge 2$ \cite{add,phen}. 
In string/M theory, $n \le 7$, and
in the brane world, $n=2$ is a reasonable choice \cite{bw}. 
In any case, $r_0$ must be fine-tuned to a very large value
$m_sr_0\sim(M_{pl}/m_s)^{2/n}\gg 1$; 
this fine-tuning problem is known as the radion problem.
In this paper, we shall simply assume that the radius at $r_0$ is a stable 
minimum. However, as we shall see, this is not the end of the radion problem.
One must still find a way for the radius to get to its final value 
without violating cosmological bounds. 
Some of the cosmological issues in the brane world have been discussed 
already \cite{cosm,az,dt,dvali,adkm,bend}.

The main concern of the present paper is the cosmology in this
framework, during the epoch before big bang nucleosynthesis, especially 
inflation \cite{guth}. 
Obviously, in the brane world, the standard cosmological 
picture is altered dramatically. 
In this paper, we present a plausible cosmological scenario where a number 
of issues in the brane world, such as inflation, density perturbation, 
reheating, baryogenesis, as well as the radion problem, are addressed. 
Our scenario incorporates the brane inflation feature \cite{dt} 
and some of its extensions \cite{dvali}, as well as some features 
of the rapid asymmetric inflation \cite{adkm}. 
The main goal here is to show that cosmology in the brane world is viable, 
and highlight some of the issues that we believe to be important.

The reader may view this scenario as a search for viable potentials 
for the inflaton and the radion. The particular form we use has an
effective potential for the
radion field $r$ and inflaton fields $\psivec = (\psi_1, \ldots,
\psi_N)$ such that 
\begin{equation}
V(\psivec,r)=V_0(\psivec)[1+\fI(r)]+f_0(r)+V_1(\psivec)
\label{eq:potent}
\end{equation}
in the Jordan frame. 
(Appendix A gives some stringy justification for such a potential.)
Here $V_0(\psivec) \sim m_s^4$ while $V_1(\psivec)
\sim m_{\rm EW}^4 \ll V_0$, where $m_{\rm EW}$ is the electroweak
scale.  Also the function $\fI(r)$ tends to force the radion to some
value $\rI$ when $V_0(\psivec)$ is large, whereas $f_0(r)$ is unimportant
until $r\to\rzero$, where $r_0$ is the value of $r$ today.   Instead
of choosing $\rzero$   
to be the only minimum of $f_0(r)$, we choose $f_0(r)$ to have
multiple minima, with $\rzero$ just one of many possible minima. 
%(In fact, for the specific example worked out in \S\ref{sec:phase2},
%$f_0(r)$ has an infinite number of minima.)
Our scenario utilizes an inflaton potential $V_0(\psivec)$ due to brane 
separation (see Appendix B), and incorporates a radion potential
$f_0(r)$ (and $\fI(r)$ as well) with multiple minima.
To simplify the problem, we might assume that $V_0(\psivec)$ depends only on 
one component of $\psivec$ and $V_1(\psivec)$ on the others (perhaps
only one other); or, $\psivec$ might only have one component with $V_0$ and
$V_1$ both depending on this one component.
The specific choices we examine for these potentials and functions are
given in Eq.\ (\ref{V0def}) for $V_0(\psivec)$, Eq.\
(\ref{fdef}) for $f_I(r)$, and Eqs.\ (\ref{f0def}), (\ref{f0defa}) and
(\ref{f0defb}) for $f_0(r)$. 
Here is a brief chronological description of the various phases of the 
scenario.

\medskip
\noindent
{\bf Phase 0: The pre-inflationary phase}:   
The key feature of this phase is that the radion field $r$ is driven
to a value $r_I$ at which it becomes fixed, thus allowing the
subsequent stage of standard inflation to take place.  The piece of
the potential (\ref{eq:potent}) that achieves this trapping is the
term $V_0(\psivec) f_I(r)$; $r_I$ will be a local minimum of
$f_I(r)$.  The initial conditions for this phase that we assume are
that the radii of the extra dimensions begin at values ``around''
$1/m_s$, where $m_s$ is the string scale (say, around 10 TeV).  
[By ``around'' we mean that values $\sim 1-100$ times larger than
$m_s^{-1}$ are not out of the question.]  Such initial conditions are
natural since in string theory, the only scale is the string scale,
and thus all parameters should typically scale like $m_s$   
unless there are good reasons (such as dynamical evolution) for other
values.  In \S\ref{sec:phase0} we explore conditions under which
the radion potential $\fI(r)$ achieves the fixing of the radion to
some value $\rI$.   A specific choice of functional form of $f_I(r)$,
which may or may not correspond to reality, is discussed further
in Appendix \ref{sec:phase0more}. Generally speaking, we think it
likely that if $f_I(r)$ has numerous potential minima separated by
some scale $\sim m^{-1}$, then the radion will become trapped at
a minimum at fairly large $mr$ (i.e. $ mr\gta$ a few), and that, if
the radion settles to a potential minimum, it does so right away,
without moving away from the potential well it starts in. 
The reason is that it is the Einstein-frame radion potential 
$\propto r^{-2 n} f_I(r)$ that is relevant to the dynamics, not the
Jordan-frame potential $\propto f_I(r)$ [see Eqs.\
(\ref{rPhirelation}) and (\ref{basic:einstein}) below,
and Appendix \ref{sec:jordein} for a discussion of the Jordan and
Einstein frames]. It is plausible that $f_I(r)$ might have fixed
amplitude of variation, so that the Einstein-frame potential will
decrease $\propto r^{-2n}$ at large $r$.  
Even under these assumptions about $\fI(r)$, the dual
conditions that $\rI$ should be ``around'' $m_s^{-1}$ and that
$m\rI\gta$ (a few), can be satisfied for values of $m$ ``around''
$m_s$. If $\fI(r)$ increases in amplitude at large 
$r$ rapidly enough to overcome the factor $r^{-2n}$, then it is possible
that $r$ actually increases somewhat from its original value during
this pre-inflation era before settling into a minimum at $\rI$.

\medskip
\noindent
{\bf Phase I: Inflation at small Planck mass:}  Once the radion is
fixed at the value $r_I$, slow-roll inflation can take place.  
The value of the Planck mass during inflation $\mplI$ is much smaller
than today's Planck mass $\mpl$, as $\mplI^2 = (r_I/r_0)^n \mpl^2$,
where $r_0$ is the value of $r$ today.
In the brane world,
brane inflation \cite{dt} is quite natural. (A brief review is given
in Appendix B.)  In the brane inflation scenario, 
when branes are separated by a distance $d$, an effective potential
$V(d)$ is generated by gravitational and other closed string exchanges
between the branes.  The distance $d$ plays the role of the inflaton.
On the other hand, $d$ is related to the vacuum expectation value of a
brane mode $\psi = m_s^2 d$, so $V(\psi)$ is a function of a brane mode.  
In a particularly intriguing scenario, the electroweak Higgs field in 
the standard model plays the role of the inflaton $\psi$. In the $n=2$
case, the inflaton potential (schematically shown in Fig. 1), 
may be taken to have the following (oversimplified) qualitative form
\begin{equation}
 V_0(\psi) + V_1(\psi) \sim m_s^4 (1 - e^{- |\psi |/m_I})+ V_1(\psi)
\label{eq:potI}
\end{equation}
with its minimum at $\psi \sim m_{EW} \sim $100 GeV.

\begin{figure}[b]
\begin{center}
\hspace{1cm}
\epsfbox{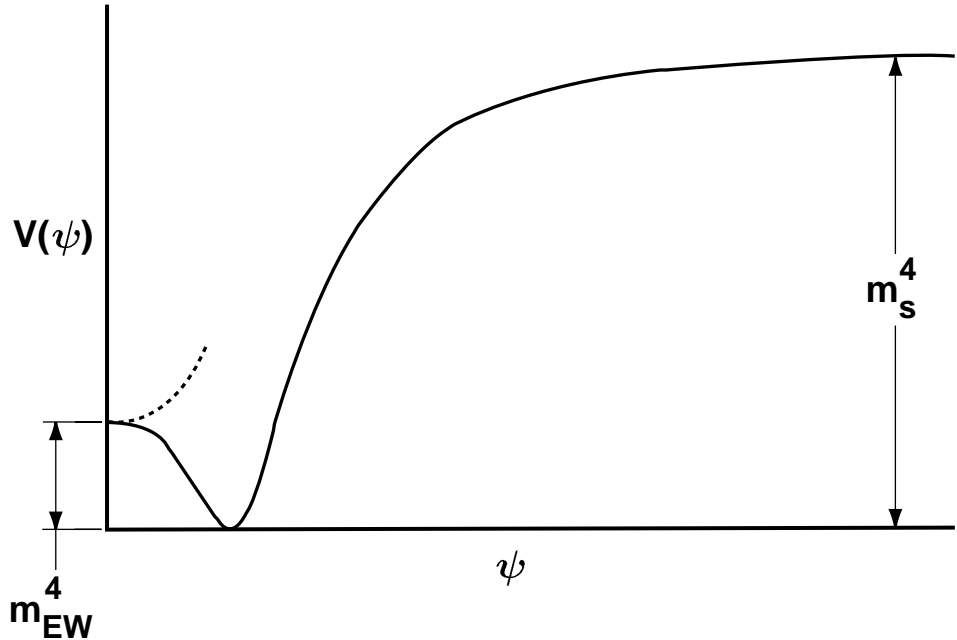}
\caption{The inflaton effective potential (not drawn to scale).
The dashed line indicates the finite temperature effective potential
after phase I.}
\end{center}
\label{fig:potential}
\end{figure}

For large $d$ (but still much smaller than $r_I$), $V \sim m_s^4$ 
is very flat, so the inflaton slowly 
rolls down the potential towards small $d$. 
In standard cosmology, the number of e-foldings required to solve 
the flatness and the horizon problems is around 60, but for brane
cosmology, the required number of e-foldings is different
[see Eqs.\ (\ref{eq:efolds}) and (\ref{eq:expconstrain})],
partly because the energy scale of inflation is $m_s\ll m_{GUT}
\sim 10^{12}$TeV, but also because the Planck mass $\mplI$ during inflation
is considerably smaller than today. [Fortuitously, the number of
e-foldings required to solve the horizon and flatness problems may
turn out to be about 60, see Eq.\ (\ref{eq:expconstrain}).]

The amplitude of primordial
density perturbations generated by quantum fluctuations
in the inflaton field during this inflationary epoch is
$\sim m_s^2/m_I\mplI$ for the inflaton potential, Eq.\ (\ref{eq:potI}).
[See Eq.\ (\ref{eq:densitypert}) in \S\ref{sec:phase0}.]
Since $\mplI\sim m_s(m_s\rI)^{n/2}$, this amplitude is approximately
$\sim m_s/m_I(m_s\rI)^{n/2}$, and to achieve the measured amplitude
$\sim 10^{-5}$ of  primordial perturbations \cite{cmbrefs}, 
we must require $(m_sr_I)^{n/2}\sim 10^5m_s/m_I$.
For $m_I\sim m_s$, this would mean that $m_sr_I\gg 1$ (e.g. 
$\sim 10^5$ for $n=2$), but larger $m_I$ is expected naively
(see Appendix \ref{sec:braneinf}), and it is conceivable that $m_s\rI
\sim 1-100$ (i.e. larger than one but not by a factor as large
as $10^5$).

\medskip
\noindent
{\bf Phase II: Radion growth and radiation domination:}  At the end of
inflation, we 
expect the brane to be heated (since  
the inflaton $\psi$ is a brane mode) while the bulk remains relatively 
cold. We expect the reheat temperature $T_r$ to be below the Hagedorn 
temperature, which is typically lower than $m_s$ (say by a factor of 3 
to 10). Since $m_s >> m_{EW}$, this temperature $T_r$ should be above 
the electroweak phase transition critical temperature $\Tc1$.
So the finite temperature effective potential $V(\psi, T_r)$ has a minimum 
at the origin, as indicated by the dashed line in Fig.\
%\ref{fig:potential}.  
1.  The inflaton rolls past the minimum of the $T=0$ potential towards
$\psi=0$ and is trapped there.  The rate of 
cooling to the bulk ($\sim T^{n+7}/m_s^{n+2}$) is very small, since $T/m_s$ 
is small. In this radiation dominated phase, the radion potential is  
negligible (which is not hard to arrange), while the inflaton field remains
frozen.  Under these conditions, we find that the
cosmological model tends toward a fixed point in which the 
temperature remains nearly constant, while the radius $r$ grows 
as a power of time [see Eqs.\ (\ref{eq:ufix}) and (\ref{eq:rhofix})].
While this powerlaw solution holds, the radion potential energy is
unimportant compared with its kinetic energy, which, however,
decreases with time (and hence with increasing radius).

After sufficient time elapses, the kinetic energy of the
radion field drops to a value comparable with the amplitude of the
radion potential $f_0(r)$, and the growth of the radion, which is
substantial 
up to this point, is halted. If $f_0(r)$ had only a single minimum,
it would be a fantastic coincidence if (a) that minimum were precisely
at the value $r_0$ and (b) the powerlaw growth of $r$ halted
exactly when that minimum was encountered. Since powerlaw growth
of $r$ during the radiation dominated phase that follows inflation
is generic in our picture, it seems that we must require that
the radion effective potential $f_0(r)$ have multiple minima.
Since it is inevitable that the radion kinetic energy becomes
smaller than the height of its effective potential after some
elapsed time, the radion must become trapped at one of the minima
of $f_0(r)$ eventually. As an illustration, we consider a periodic
radion potential; $r_0$ is just one of the infinity of minima
of this potential. That the Universe settles to $r_0$ is
a cosmological accident in our scenario, although it is natural
for the radion to settle to some radius much larger than
its value during inflation (and much larger than the string
scale $m_s^{-1}$). Thus, the radion problem is not so severe
in our picture, which accommodates growth of the radius
of the extra dimensions to a large, but stable value very simply.
What we do not explain is why $r_0$ (and hence Newton's
constant) has a particular value among the infinity of
possibilities.
However, we identify what conditions must be satisfied by the
underlying physical theory for the Universe to settle at
$r_0$ [see Eqs.\ (\ref{eq:ubulkrzn}) and (\ref{eq:ubulkrzn1})].

\medskip
\noindent
{\bf Phase III:  Second inflationary era and electroweak phase
transition:}  When $T$ drops below $\Tc1$, the stable minimum of the
inflaton will yield the 
spontaneous symmetry breaking of the electroweak model. 
It is reasonable to suppose that the electroweak phase transition is 
first-order. (This should be easy to arrange in models with multi-Higgs 
fields, {\em e.g.}, the minimal supersymmetric standard model.) In this case, 
some super-cooling is expected and the actual phase transition happens 
during a period some time after $T$ has dropped below $\Tc1$. 
In the mean time, we expect considerable dilution of the radion energy density
as well as the bulk energy density. Indeed, we show that requiring the
radion density at present not to exceed the critical density for a flat
Universe constrains the number of e-foldings of this inflationary 
era [see Eqs.\ (\ref{eq:omegaradion}) and (\ref{eq:f2lim})].
[Associated with the radion density today would be small-amplitude
oscillations of Newton's constant at a high frequency \cite{accstein},
\cite{swill}; see Eqs.\ (\ref{eq:radionmass}) and \S\ref{sec:discussion}.]
This can be achieved by a short period of inflation during the 
super-cooling period, followed by either prompt or delayed reheating.
The actual electroweak phase transition then takes place with the presence 
of nucleation bubbles.
This allows the electroweak phase transition to 
complete and baryogenesis during this period can happen more or less as 
in the standard scenario \cite{baryo}. Alternatively, baryogenesis can 
happen via the Affleck-Dine mechanism or some other mechanism \cite{ad}.
In fact, it may take place before the end of the second inflationary era.

The final reheat temperature can be around a few GeV, maybe even close 
to the electroweak scale $m_{EW}$, if $m_s$ is large enough, and still
avoid excessive cooling to the bulk, which would overproduce Kaluza-Klein
(KK) modes, whose energy density could overclose the Universe and
ruin the success of Big Bang nucleosynthesis.
It is also high enough to provide the "initial conditions" of the
hot Big Bang before Big Bang nucleosynthesis.

The scenario is summarized in Fig. 2.

The basic plan of this paper is the following. In \S\ref{sec:matching},
we present our cosmological scenario; \S\ref{sec:setup} gives some
useful background (some of which is also found in Appendix \ref{sec:jordein}),
\S\ref{sec:phase0}
treats the pre-inflationary phase, \S\ref{sec:phase1} treats inflation at
small Planck mass, and \S\ref{sec:phase2} treats the phase during which
the radius of the extra dimensions grows from $\rI$ to $r_0$. Some
constraints on our model are gathered in \S\ref{sec:exprad}. Density
fluctuations during the epoch when the radion grows are discussed 
briefly in \S\ref{sec:fluct}. The results are discussed in
\S\ref{sec:discussion}. Some additional details about our cosmological
model are contained in various Appendices.

\section{Matching Phases}
\label{sec:matching}

\subsection{Setup}
\label{sec:setup}

The starting point for our analysis is the following low energy action,
which is valid when the lengthscales over which all fields vary are
much larger than the size of the extra dimensions:
\begin{eqnarray}
S &=& \int d^4 x \sqrt{- {\hat g}} \bigg[ {{\hat R} \over 16 \pi
G } - {1 \over 2} ({\hat \nabla} \Phi)^2 
%\nonumber \\ \mbox{} && 
- {1 \over 2} e^{-\Phi/\mu} ({\hat \nabla} \psivec)^2 -
e^{-2 \Phi/\mu}\, V(\psivec, \Phi) \bigg]
\nonumber \\
\mbox{} && + S_{\rm rest}[e^{-\Phi/\mu} {\hat g}_{\alpha\beta},
\chi_{\rm rest}].
\label{action2}
\end{eqnarray}
This action is derived from the higher dimensional description in
Appendix \ref{sec:jordein}.  
The action is written in the Einstein frame,
${\hat g}_{\alpha\beta}$ is the Einstein frame metric, and 
\beq
g_{\alpha\beta} = e^{-\Phi/\mu} {\hat g}_{\alpha\beta}
\eeq
is the physical, Jordan frame metric.  The quantity $G$ is the usual
3-dimensional Newton's constant, and $\mu$ is a mass of
order the Planck mass $\mpl = \sqrt{c \hbar / G} = 1.22 \times
10^{19} \, {\rm GeV}$ given by
\beq
\mu = \mpl \,  \sqrt{ {n+2 \over 32 \pi n} },
\label{mudef}
\eeq
where $n$ is the number of extra dimensions.  The field $\Phi$ is the
canonically normalized radion field, related to the radius $r$ of the
extra dimensions by 
\beq
r = r_0 \exp \left[{\Phi \over  n \mu }\right],
\eeq
where $r_0$ is the equilibrium radius of the extra dimensions today.
The fields $\psivec = (\psi_1, \ldots, \psi_N)$ are inflaton fields.
The quantity $V(\psivec,\Phi)$ is the Jordan-frame potential for the
radion and inflaton discussed
in the Introduction and in Appendix \ref{sec:effpot} below.  Finally
the action $S_{\rm rest}[g_{\alpha\beta},\chi_{\rm rest}]$ is the
action of the remaining matter fields $\chi_{\rm rest}$, which in our
analysis below we will treat as a fluid.

As discussed above, we assume that the Jordan-frame effective
potential $V(\psivec,\Phi) = V(\psivec,r)$ is of the form [cf.\ Eq.\
(\ref{eq:potent})]
\be
V(\psivec,r)=V_0(\psivec)[1+\fI(r)]+f_0(r)+V_1(\psivec).
\ee
Here we might assume that $V_0$ depends only on one component
of $\psivec$ and $V_1$ on the other components (perhaps only one other).
The function $\fI(r)$ tends to force the radion to some value
$\rI$ while $V_0$ is large, whereas $f_0(r)$ is unimportant
until $r\to\rzero$, its value today. Following Ref.\ \cite{dt}, 
we assume that $V_0(\psivec)\to 0$ as $\psivec\to 0$
and that $V_0(\psivec)$ asymptotes to a constant value, $\Vhatz$, 
exponentially with some mass scale(s) $\mI$. The potential
$V_1(\psivec)$ is assumed to have a minimum at nonzero
$\psivec$, and a value $\Vhat1 \equiv V_1(0)$ at $\psivec=0$ which
satisfies $\Vhat1 \ll \Vhatz$.

We will do all our calculations in the Einstein frame.  
Assuming zero spatial
curvature, the metric of the cosmological background in the Jordan
frame can be written in the form
\be
ds^2=-dt^2+a^2(t)d\xvec\cdotvec d\xvec,
\label{eq:jbkgnd}
\ee
and the corresponding metric in the Einstein frame is
\be
d\sb^2= e^{\Phi/\mu} \left[ -dt^2+a^2(t)d\xvec\cdotvec d\xvec \right] = 
-d\tb^2+\ab^2(\tb)d\xvec\cdotvec d\xvec,
\label{eq:ebkgnd}
\ee
where 
\be
d\tb=\exp(\Phi/2\mu)\ dt\qquad\qquad\ab(\tb)=\exp(\Phi/2\mu)
\ a(t).
\label{eq:scalings}
\ee
It is important to keep the relations (\ref{eq:scalings}) in mind,
since proper time is not the same in the Einstein and Jordan frames,
nor is the scale factor. In terms of $r$, the scalings are
\be
d\tb=(r/\rzero)^{n/2}dt\qquad\qquad\ab(\tb)=(r/\rzero)^{n/2}
a(t).
\ee
Thus, as $r$ changes, the relative rates of advance of time and scale factor
differ in the two frames.

We treat the last term in the action (\ref{action2}) as a fluid with
Jordan-frame density $\rho$ and pressure $p$.  Then, the
cosmological equations of motion that follow from the action
(\ref{action2}) follow from the general equations of motion given in
Appendix \ref{sec:jordein}, and are given by \footnote{These evolution
equations do not include any coupling between the inflaton and the
radiation, which would be necessary to describe reheating.  If we add
the standard type of phenomenological terms to achieve this in the Jordan
frame, we obtain after transforming to the Einstein frame that one
should add a term $-F(\psi) \psi^\prime e^{-\Phi/(2 \mu)}$ to the
right hand side of the inflaton equation (\ref{basic:inflaton}), and a
term $F(\psi) \psi^{\prime 2} e^{\Phi / (2 \mu)}$ to the right hand
side of Eq.\ (\ref{basic:fluid}), where $F(\psi)$ can be any function.}
\beq
{\hat H}^2 = {8 \pi  \over 3 \mpl^2} \bigg[ {1
\over 2} {\Phi}^{\prime 2} + {1 \over 2} e^{-\Phi/\mu} {\psi}^{\prime 2}
%\nonumber \\ \mbox{} && 
+ e^{-2 \Phi/\mu} V(\psi,\Phi) 
+ e^{- 2 \Phi / \mu} \rho \bigg],
\label{basic:einstein}
\eeq
\begin{eqnarray}
{\Phi}^{\prime\prime} + 3 {\hat H} {\Phi}^\prime + &&{\partial \over
\partial \Phi} \left[ e^{- 2 \Phi/\mu}  V(\psi,\Phi)\right]
+ {1 \over 2 \mu} e^{-\Phi/\mu} \psi^{\prime 2} 
%\nonumber \\ \mbox{} &&
= { \rho - 3 p \over 2 \mu} e^{- 2 \Phi / \mu},
\label{basic:radion}
\end{eqnarray}
\beq
{\psi}^{\prime\prime} + {\psi}^\prime \left[ 3 {\hat H} - {\Phi^\prime
\over \mu} \right] + e^{-\Phi/\mu} {\partial \over \partial \psi}
V(\psi,\Phi)  =0.
\label{basic:inflaton}
\eeq
and
\beq
{\rho}^\prime + 3 (\rho + p) \left[ {\hat H} - {1 \over 2 \mu}
{\Phi}^\prime \right]=0.  
\label{basic:fluid}
\eeq
In these equations primes denote derivatives with respect to
Einstein-frame proper time ${\hat t}$ [Eq.\ (\ref{eq:jbkgnd}) above], and
${\hat H}$ is the Einstein-frame Hubble parameter ${\hat H} = {\hat
a}^\prime / {\hat a}$.

In the next few subsections, we shall substitute the potential
(\ref{eq:potent}) into the evolution equations (\ref{basic:einstein}) --
(\ref{basic:fluid}), and solve for approximate solutions in the four
different phases of cosmological evolution discussed in the
Introduction.

\subsection{Phase 0: Radion to Its First Equilibrium}
\label{sec:phase0}

During Phase 0, the radion evolves to some size $\rI$ where
it remains pinned during inflation (phase I). This pinning happens
by some time $\tb_0$ when the scale factor is $\ab_0$. The
end of phase 0 signals the onset of the first inflationary
phase, phase I

To see how the pinning might come about, let us assume that we can
neglect the terms $f_0(r)$ and $V_1(\psivec)$, and that there is no 
energy density except what is due to $\psivec$ and the radion.
Furthermore, let us assume that the kinetic
energy of the inflaton is negligible, and that $V_0(\psivec)\approx
\Vhatz$. Then the Friedmann equations (\ref{basic:einstein}) and
(\ref{basic:radion}) together with the potential (\ref{eq:potent})
simplify to 
\baray
\Hbar^2={8\pi\over 3\mpl^2}\biggl\{\Vhatz\exp(-2\Phi/\mu)
[1+\fI(r)]
+{1\over 2}(\Phi^\prime)^2\biggr\}
\nonumber\\
\Phi^{\prime\prime}+3\Hbar\Phi^\prime=
{2\Vhatz\exp(-2\Phi/\mu)\over\mu}
\biggl[1+\fI(r)-{r\over 2n}{d\fI(r)\over dr}\biggr].
\label{F0}
\earay
We can render these equations non-dimensional by letting
\be
\zeta=\exp[2(\Phi-\Phi_i)/\mu]
\ee
and defining a new time variable $\tau$ by
\be
d\taub=\biggl({8\pi {\hat V}_0 \over
3\mpl^2}\biggr)^{1/2}\exp(-\Phi_i/\mu)d\tb, 
\ee
where $\Phi_i$ is the initial value of $\Phi$, at the beginning
of Phase 0. In terms of these new variables, we find
($f^\prime=df/d\taub$ in these equations)
\baray
\biggl({y^\prime\over y}\biggr)^2=
{1+\fI(r)\over\zeta}+{1\over 2\nu}\biggl({\zeta^\prime\over
\zeta}\biggr)^2\nonumber\\
\zeta^{\prime\prime}+\biggl({3y^\prime\over y}-{\zeta^\prime\over\zeta}
\biggr)\zeta^\prime=\nu\biggl[1+\fI(r)-{r\over 2n}{d\fI(r)\over dr}
\biggr].
\label{eq:phase01}
\earay
Here $y=\ab/\ab_i$, with $\ab_i$ the initial value of the Einstein
frame scale factor, $r=r_i\zeta^{1/2n}$, with $r_i$ the initial
value of the radius of the extra dimensions, and
\be
\nu\equiv{48n\over n+2}.
\ee
It is easy to see that if $\fI(r)=0$, the solution of Eq. (\ref{eq:phase01})
tends to 
\be
y\propto\taub^{1/3}\qquad\zeta\propto\taub^{\sqrt{2\nu}/3},
\ee
according to which $r$ grows without bound, and the radion kinetic
energy dominates the energy density of the Universe, but there is
no inflation \cite{berkin}. The approach
to this asymptotic solution may be very slow: for example, for $n=2$
we have $\zeta\propto\taub^{4/\sqrt{3}}$, so the vacuum energy 
density declines $\propto\taub^{4/\sqrt{3}}$, which is only a bit
faster than the rate of decline of the radion kinetic energy,
$1/\taub^2$. Nevertheless, it is noteworthy that without the
radion potential, $r$ would grow to infinity in this phase.

To halt the growth of $r$, the radion potential must be capable
of trapping the radion field. From Eq.\ (\ref{eq:phase01}), we see that
this is only possible if the condition 
\be
1+\fI(r)-{r\over 2n}{d\fI(r)\over dr}=0
\label{eq:stopcond}
\ee
can be satisfied.  If we consider the choice $f_I(r) = (m r)^\kappa$
for example, where $m$ is a mass parameter and $\kappa > 0$, then 
it is clear 
that Eq.\ (\ref{eq:stopcond})
will only have real solutions for $\kappa>2n$ in which case the radion
could settle to a value $mr=(\kappa/2n-1)^{-1}$.  (Analogously shifted
minima have been discussed by e.g. Steinhardt \& Will \cite{swill}.) 
However, such steeply growing functions $f(r)$ could prevent $r$ from
growing to a large value later on.

Instead, we shall consider the possibility that 
\be
\fI(r)=aF(r),
\ee
where $a$ is a dimensionless amplitude factor, and $F(r)$ has multiple
minima, separated by a characteristic scale $\sim m^{-1}$, with ``potential
barriers'' $\vert F(r)\vert\sim 1$. A specific example (but not unique
or required) is $F(r)=1-\cos mr$, for which Eq.\ (\ref{eq:stopcond})
becomes
\be
1+a(1-\cos mr)-{amr\sin mr\over 2n}=0.
\ee
Since $\cos mr\leq 1$ and $\sin mr\leq 1$, there are no solutions to this
equation unless $amr/2n>1$, or $mr>2n/a$. Thus, unless $a$ is large (which
we consider unlikely), the radion will only settle into minima at relatively
large values of $mr$, if at all. This conclusion ought to hold for
other choices of $F(r)$ with similar qualitative properties. If, for
example, $m\sim m_s$, then we conclude that the radion will only settle
on values larger than the string scale, which, in fact, is required for
consistency of our entire picture. (Remember that, for example, $r$
must exceed the brane separation.)

For $\fI(r)$ of this general type, it seems likely that the radion must
settle into its first minimum, the one nearest its value at the onset
of Phase 0, if it settles to a minimum at all. The reason is that
the height of the Einstein frame effective potential decreases with
increasing $r$, so that if the radion acquires sufficient kinetic
energy to roll over the first barrier it encounters, it should
be able to overcome all subsequent barriers. (Remember that in the
asymptotic solutions to Eq.\ [\ref{eq:phase01}] the energy density of the
Universe becomes dominated by radion kinetic energy as time progresses.)
We explore the particular example $F(r)=1-\cos mr$ in some detail
in Appendix \ref{sec:phase0more}. It is also important to note, though, that
it is possible to imagine choices for $\fI(r)$ that undo the decrease
factor in amplitude $\propto r^{-2n}$ in Eq.\ (\ref{F0}) at large
$r$. For such potentials, it 
might be possible for $r$ to evolve considerably before settling to
a potential minimum.

\subsection{Phase I: Inflation at Small Planck Mass}
\label{sec:phase1}

During Phase I, the Universe inflates at a fixed radion 
radius, $\rI$. This fixes the Planck mass to be 
\be
\mplI^2=(\rI/\rzero)^n\mpl^2,
\ee
and the expansion rate is
\be
\Hbar_I^2=\biggl({\ab^\prime\over\ab}\biggr)^2
={8\pi\rzero^{2n}\Vhatz\over 3\mpl^2\rI^{2n}}
={8\pi\rzero^n\Vhatz\over 3\mplI^2\rI^n}
\ee
in the Einstein frame, where prime denotes differentiation
with respect to $\tb$. The Jordan frame expansion rate
is simply
\be
H_I^2=\biggl({\dot a\over a}\biggr)^2=
{8\pi \Vhatz\over 3\mplI^2},
\ee
where dot denotes differentiation with respect to $t$; this
is just as in the usual general relativity but with a different
Planck mass. The scale factor in this phase is
\be
{\ab\over\ab_0}=\exp[\Hbar_I(\tb-\tb_0)]={a\over a_0}
=\exp[H_I(t-t_0)].
\ee
Since the radius of the extra dimensions is frozen in phase I, it is
equally easy to use the Jordan or Einstein frame descriptions. (The
same will not be true for subsequent phases.)

Presuming that a single inflaton field is important in this
phase, with an effective potential
\be
V_0(\psi)=\Vhatz[1-\exp(-\psi/\mI)],
\label{V0def}
\ee
the inflaton equation of motion is
\be
\ddot\psi+3H_I\dot\psi=-{\Vhatz\over\mI}\exp(-\psi/\mI).
\ee
The evolution of the inflaton proceeds in two subphases. During
the first subphase, we have the usual slow-rolling approximation,
\be
\dot\psi\approx -{\Vhatz\exp(-\psi/\mI)\over 3H_I\mI},
\ee
whose solution is
\be
\exp(\psi/\mI)=\exp(\psi_0/\mI)-{\Vhatz(t-t_0)\over 3\mI^2H_I}
=\exp(\psi_0/\mI)-{\mplI^2\over 8\pi\mI^2}H_I(t-t_0).
\ee
This approximate solution holds as long as the two conditions
\be
\biggl\vert{\ddot\psi\over 3H_I \dot\psi}\biggr\vert
\approx{\mplI^2\exp(-\psi/\mI)\over 24\pi\mI^2}\ll 1
\qquad\qquad
{\dot\psi^2\over 2\Vhatz}\approx 
{\mplI^2\exp(-2\psi/\mI)\over 48\pi\mI^2}\ll 1
\ee
are satisfied. Assuming that $\mplI\gg\mI$, which emerges
naturally later, the first of these two conditions fails
first, when 
\be
\exp(\psi/\mI)\sim {\mplI^2\over 24\pi\mI^2}.
\label{lim}
\ee
If the initial value $\exp(\psi_0/\mI)$ of $\exp(\psi/\mI)$ is far
larger than the limit (\ref{lim}), then the
time required for the inflaton field to reach this 
magnitude is extremely large, given by\footnote{The subscript
``sr'' stands for the end of slow rolling.}
\be
H_I(t_{\rm sr}-t_0)\approx {8\pi\mI^2\over\mplI^2}\exp(\psi_0/\mI),
\ee
where $t_{\rm sr}$ is the time at the end of slow-roll, which implies many
e-foldings during inflation.  When slow-rolling 
ends, a second sub-phase of inflaton evolution begins. Since
the kinetic energy of the inflaton at the beginning of
this subphase is only
$\sim (12\pi\mI^2/\mplI^2)\Vhatz\ll\Vhatz$, the inflaton
moves on approximately a ``zero energy solution with negligible
damping''.  That is, it satisfies the equation
\be
\dot\psi\approx -\sqrt{2\Vhatz}\exp(-\psi/2\mI),
\ee
which has the solution
\be
\exp(\psi/2\mI)=\exp(\psi_{sr}/2\mI)-\sqrt{\Vhatz\over 2\mI^2}
(t-t_{\rm sr})
=\exp(\psi_{sr}/2\mI)-\sqrt{3\over 16\pi}{\mplI\over\mI}
H_I(t-t_{\rm sr}),
\ee
where $\exp(\psi_{sr}/2\mI)\sim\mplI/\mI\sqrt{24\pi}$. The
time remaining for $\psi\to 0$ is not large: $H_I (t_{\rm end}-t_{\rm sr})
\sim 1$, where $t_{\rm end}$ signifies the end of this inflationary
epoch, and henceforth we do not distinguish between the
two times $t_{\rm sr}$ and $t_{\rm end}$.

Using our approximate expression for $H_I (t_{\rm sr}-t_0)\approx
H_I (t_{\rm end}-t_0)$, we rewrite the slow rolling solution as
\be
\exp(\psi/\mI)\approx{\mplI^2\over 8\pi\mI^2}H_I (t_{\rm end}-t).
\ee
{}From this and the slow rolling approximation it follows
that 
\be
\dot\psi\approx -{\mI\over t_{\rm end}-t}.
\ee
The primordial density perturbation amplitude is then
\cite{densityperts}
\be
{\delta\rho\over\rho}\sim {H_I^2\over\dot\psi}
\approx{\sqrt{8\pi\Vhatz/3}\over\mI\mplI}
N_k,
\label{eq:densitypert}
\ee
where $N_k$ is the number of e-foldings that remain between
horizon crossing for a scale of comoving length $\sim k^{-1}$
and the end of Phase I. We note that the spectrum of inhomogeneities
implied by Eq.\ (\ref{eq:densitypert}) is insensitive to $k$,
in agreement with observations \cite{cmbrefs}.
If the effective potential as a function of
interbrane separation $d$ is proportional to $ 1-\exp(-m_dd)$, and we
set $\psi=m_s^2d$, then $m_dd=m_d\psi/m_s^2$, which implies
$\mI=m_s^2/m_d$. Then the amplitude of the density fluctuations
is proportional to $\sqrt{\Vhatz}m_d/m_s^2\mplI$, which,
for $\Vhatz\sim m_s^4$, is $\sim m_d/\mplI$.
Note also that since the radion is trapped, fluctuations in $r$ are
suppressed [see Eq.\ (\ref{eq:iralm}) of Appendix \ref{sec:phase0more}].

\subsection{Phase II: Radiation Domination}
\label{sec:phase2}

At the end of inflation, $\tb=\tb_1$ and $\ab=\ab_1$; the
Universe reheats to a temperature $\epsilon\Vhatz^{1/4}$
where $\epsilon<1$ depends on how efficiently the kinetic
energy of the inflaton is thermalized once $\psivec\to 0$.

Reheating will alter the effective potential so that there
is a minimum at $\psivec=0$, provided that the critical 
temperature $\Tc1$ for the phase transition (presumed first order)
connected with the potential $V_1(\psivec)$ is small compared
with the reheating temperature. In this case, there will
be a nonzero vacuum energy $\Vhat1$, but as long as the
temperature remains above $\Tc1$, the Universe remains
radiation-dominated, and $\psivec$ is pinned at zero.

We assume that the radion potential becomes negligible
when this happens. Let us explore the growth in the
radion field that ensues.

It is most convenient to work in the Einstein frame.
This phase of evolution will divide into three subphases.
During the first subphase,
when radiation dominates, the Friedmann equations are [cf.\ Eqs.\
(\ref{basic:einstein}) and (\ref{basic:fluid}) above with $p = \rho/3$]
\be
\Hbar^2={8\pi\over 3\mpl^2}\rho\exp(-2\Phi/\mu)
\qquad\qquad
\rho\ab^4\exp(-2\Phi/\mu)={\rm constant},
\ee
where we recall that $\rho$ is the {\it Jordan} frame
energy density i.e. $\rho\sim T^4$. Thus, the scale
factor in Einstein frame is simply
\be
\ab=\ab_1\biggl({\tb\over\tb_1}\biggr)^{1/2},
\label{asoln}
\ee
just as it would be in constant-$\mpl$ cosmology.
However, to find out how fast the temperature decreases,
we also need to know how the radion field evolves.

Under the assumption that the radion potential is
negligible, we find from Eqs.\ (\ref{eq:potent}), (\ref{basic:radion}) and
(\ref{V0def}) the evolution equation
\be
\Phi^{\prime\prime}+3\Hbar\Phi^\prime
={2\Vhat1\over\mu}\exp(-2\Phi/\mu).
\label{eq:radion1}
\ee
We make the change of variables
\beq
u=\exp(2\Phi/\mu).
\label{udef}
\eeq
Then Eq.\ (\ref{eq:radion1})
is equivalent to
\be
u^{\prime\prime}+\biggl({3\Hbar}-{u^\prime\over u}
\biggr)u^\prime
=u^{\prime\prime}+\biggl({3\over 2\tb}-{u^\prime\over u}
\biggr)u^\prime
={4\Vhat1\over\mu^2},
\label{eq:uevolv}
\ee
where $\Hbar=1/2\tb$, appropriate for this subphase,
has been used. During the first subphase, 
\be
{3\over 2\tb}\gg{u^\prime\over u}.
\label{eq:subphase1}
\ee
In that case, under the assumption that $u^\prime=0$
and $u=u_1$ at $\tb=\tb_1$, we find
\be
u^\prime={8\Vhat1\over 5\mu^2}\biggl(\tb-{\tb_1^{5/2}
\over\tb^{3/2}}\biggr),
\ee
which can be integrated to yield
\be
u=u_1+{4\Vhat1\over 5\mu^2}
\biggl(\tb^2-5\tb_1^2+{4\tb_1^{5/2}\over\tb^{1/2}}
\biggr).
\ee
This solution holds as long as Eq.\ (\ref{eq:subphase1})
is true, a condition that fails when
\be
{2\tb u^\prime\over 3u}\approx{16\Vhat1\tb^2/15\mu^2\over
u_1+4\Vhat1\tb^2/5\mu^2}\sim 1,
\ee
or
\be
u_1\sim{\Vhat1\tb^2\over\mu^2}.
\label{eq:endofsubphase1}
\ee
(In getting this condition, we have assumed that the
first subphase ends at a time $\gg\tb_1$.)
Since $u=\exp(2\Phi/\mu)$ is still only a factor of
two or so different from $u_1$, it follows that
\be
\rho={\rho_1\ab_1^4\exp(2\Phi/\mu)\over\ab^4\exp(2\Phi_1/\mu)}
={\rho_1\Vhat1\tb_1^2u\over\mu^2u_1^2}\sim
{\rho_1\Vhat1\tb_1^2\over\mu^2u_1},
\ee
where in the first equality we used Eqs.\ (\ref{asoln}), (\ref{udef})
and (\ref{eq:endofsubphase1}).
Using\footnote{Recall that $\mplI^2=\mpl^2(\rI/\rzero)^n=\mpl^2\sqrt{u_1}
$.}
\be
t_1^2\sim{3\mplI^2\over 8\pi\rho_1}\sim
{\mpl^2\sqrt{u_1}\over\rho_1},
\ee
and $\tb_1^2=\sqrt{u_1}t_1^2,$ we find
\be
\rho\sim\Vhat1
\ee
since $\mu\sim\mpl$. Thus, the first subphase ends when the
energy density in radiation becomes comparable to the vacuum
energy in the inflaton. The value of the Planck mass only
changes by a factor of order unity during this regime.

During the second subphase, the radion potential is still
unimportant, but the radion kinetic energy is not. Thus, the
radion evolves according to Eq.\ (\ref{eq:uevolv}), but the
expansion rate in the Einstein frame is given by
\be
\Hbar^2={8\pi\over 3\mpl^2}\biggl[{\rho\over u}
+{V_1(\psi)\over u}+{\mu^2\over 8}\biggl({u^\prime\over u}
\biggr)^2
+{(\psi^\prime)^2\over 2\sqrt{u}}
\biggr].
\ee
The evolution of the inflaton field is governed by the
equation
\be
\psi^{\prime\prime}+3\Hbar\psi^\prime=
-u^{-1/2}{dV_1(\psi)\over d\psi}.
\ee
We assume that the Universe remains hot enough that
the inflaton is trapped in a symmetric phase, at fixed
fixed vacuum energy density $\Vhat1$ during this
subphase.
At least at first, the kinetic
energy of the inflaton field will be unimportant,
and we can approximate the expansion rate by
\be
\Hbar^2={8\pi\over 3\mpl^2}\biggl[{\rho\over u}
+{\Vhat1\over u}+{\mu^2(u^\prime)^2\over 8u^2}\biggr].
\label{eq:Hbar1}
\ee
It is easy to see that at the start of this subphase,
the three contributions to the energy density 
are comparable to one another. Moreover, there is a
simple, powerlaw solution to the fully nonlinear
problem defined by eqs. (\ref{eq:uevolv}) and
(\ref{eq:Hbar1}). For this solution, $\ab\propto
\sqrt{\tb}$, and 
\be
u={4\Vhat1\tb^2\over\mu^2}
\qquad{u^\prime\over u}={2\over\tb};
\label{eq:ufix}
\ee
since $\rho\ab^4/u=$constant, $\rho=$constant.
{}From Eq.\ (\ref{eq:Hbar1})
we find a consistency condition
\be
{\rho\over\Vhat1}={9n-6\over n+2},
\label{eq:rhofix}
\ee
which gives $\rho/\Vhat1=3$ for $n=2$, for example;
thus, the temperature remains $\sim\Vhat1^{1/4}$,
and could be comfortably above the critical temperature
for the inflaton in this regime.\footnote{If we let
$\ab(t)=\ab_0(t)[1+\alpha(\tb)]$ and 
$u(\tb)=u_0(\tb)[1+\eta(\tb)]$, where $\ab_0(\tb)$
and $u_0(\tb)$ are the powerlaw solution of 
eqs. (\ref{eq:uevolv}) and (\ref{eq:Hbar1}),
it is easy to show that 
$$\eta^{\prime\prime}+{3\eta^\prime\over 2\tb}
+{\eta\over\tb^2}+{6\alpha^\prime\over\tb}=0$$
$$\alpha^\prime+{\alpha\over\tau}\biggl({3n-2\over 4n}\biggr)
=\biggl({n+2\over 24n}\biggr)\biggl(\eta^\prime-
{\eta\over 2\tb}\biggr);$$
these coupled perturbation equations have powerlaw solutions
$\propto\tb^s$ where
$$(s+1)\biggl(s^2+{s\over 2}+{3n-2\over 4n}\biggl)=0.$$
Thus, there are no growing perturbations, and the powerlaw,
fixed point solution to eqs. (\ref{eq:uevolv}) and (\ref{eq:Hbar1})
is stable.}
Within this solution, it also follows that,
up to a possible additive constant,
\be
t={\sqrt{2\mu\tb}\over\Vhat1^{1/4}},
\ee
and therefore, as a function of $t$,
the radion expands according to
\be
u={\Vhat1^4t^4\over\mu^4},
\ee
and grows without any expansion of the Universe
at all in the Jordan frame.

It is worth investigating the meaning of this
powerlaw solution further. Clearly, a solution in which the
temperature remains constant is not expanding in the
Jordan frame at all. Such a solution ought to apply
only in a limiting sense. That is, the correct solution
might approach this one asymptotically, at late times.
To see if this happens, we consider the numerical solution
eqs. (\ref{eq:uevolv}) and (\ref{eq:Hbar1}), with the
energy-conservation condition $\rho\ab^4/u$=constant.
We arbitrarily choose $u=u_i$ and $\rho=\rho_i$ at some
initial time $\tb_i$, when $\ab\equiv\ab_i$. If we
define 
\be
y\equiv{\ab\over\ab_i}\qquad{\rm and}\qquad
\zeta\equiv{u\over u_i},
\ee
and define a dimensionless Einstein-frame time variable by
\be
d\taub=(8\pi\rho_i/3\mpl^2u_i)^{1/2}d\tb,
\ee
then we find the two coupled equations
\baray
\biggl({y^\prime\over y}\biggr)^2=
{1\over 2\nu}\biggl({\zeta^\prime\over\zeta}
\biggr)^2+{v\over\zeta}+{1\over y^4}\nonumber\\
\zeta^{\prime\prime}+
\biggl({3y^\prime\over y}-{\zeta^\prime\over\zeta}
\biggr)\zeta^\prime=\nu v,
\label{eq:numsolve}
\earay
where
\be
\nu\equiv{48n\over n+2}\qquad{\rm and}\qquad
v\equiv{\Vhat1\over\rho_i}.
\ee
We can also evaluate the Jordan frame scale factor from
\be
\ab=u^{1/4}a,
\ee
and can find the time elapsed in the Jordan frame using
\be
d\tb=u^{1/4}dt,
\ee 
which we rescale by defining $\tau_J=(8\pi\rho_i/3\mpl^2\sqrt{u_i})^{1/2}
t$, to get $\tau_J^\prime=\zeta^{-1/4}$. Eqs. (\ref{eq:numsolve}) are
to be solved with initial conditions $\zeta=y=1$ at $\taub=\taub_i$; we
can choose $\tb_i$ arbitrarily (although we expect it to be $\approx
0.5$ if we want a solution that has $y\to 0$ at $\taub\to 0$).
The solutions depend only on the two parameters, $\nu$ and $v$; in
terms of these, the powerlaw solution found previously becomes
\be
\zeta=\nu v\taub^2={48n v\taub^2\over n+2}\qquad\qquad
y=\biggl({1\over 4}-{3\over\nu}\biggr)^{-1/4}\taub^{1/2}
=\biggl({16n\over 3n-2}\biggr)^{1/4}\taub^{1/2}
\ee
Numerical evaluations for $n=2$ show that, 
although the solutions oscillate slightly
at late times, they approach the simple solution found above
to high accuracy. From examining the output, it appears that
the Jordan frame scale factor actually does not remain precisely
constant at late times, but increases and even decreases slightly
(by less than 10\%) as time progresses.

In the asymptotic regime of the second subphase, 
the radion grows according to
\be
u=(r/\rzero)^{2n}={4\Vhat1\tb^2\over\mu^2},
\ee
so $u\to 1$, or $r\to\rzero$, at a time
\be
\tb_{r\to\rzero}={\mu\over 2\sqrt{\Vhat1}}
\ee
if this solution continues to hold. Note that 
$\tb_{r\to\rzero}\sim u_1^{-1/2}\tb_i$, where $\tb_i$ marks
the onset of this subphase (or the end of the first subphase;
e.g. Eq.\ [\ref{eq:endofsubphase1}]); since $u_1\ll 1$,
$\tb_{r\to\rzero}\gg\tb_i$.
\footnote{As this is a pretty odd solution, let us also
consider an alternative, that the radion evolves like a free
field after the second subphase begins, and its kinetic energy
dominates the energy density of the Universe. In this case
$\Phi^\prime=\Phi^\prime_i\ab_i^3/\ab^3$, and
$$
\Hbar^2={4\pi(\Phi^\prime_i)^2\over\mpl^2(\ab/\ab_i)^6},
$$
which implies the solution
$(\ab/\ab_i)^3=\tb\sqrt{12\pi}\Phi^\prime_i/\mpl$. Using this
solution, we find that $\Phi^\prime=\mpl/\tb\sqrt{12\pi}$ and
consequently
$$
r=r_i\biggl({\tb\over\tb_i}\biggr)^{\sqrt{8/3n(n+2)}},
$$
or $r\propto \tb^{1/\sqrt{3}}\propto\ab^{\sqrt{3}}$ for
$n=2$.
For this solution to hold true, all other contributions
to the energy density must decline more rapidly than
$(\Phi^\prime)^2/2\propto\tb^{-2}$. But it is easy to
see that $\rho\exp(-2\Phi/\mu)\propto\ab^{-4}\propto
\tb^{-4/3}$ according to this solution, so it must not
be valid.}

In order for the radion to become pinned at $r=\rzero$
we need a coincidence to happen: near the time 
$\tb_{r\to\rzero}$, the radion potential itself must 
begin to play a central role in the evolution of the field.
Only the radion's potential can make it settle
into a minimum, rather than rolling forever to ever
increasing radius. Indeed, what we want is for the effective
potential of the radion to have many possible minima, so
that the value it settles into eventually is determined by
this coincidence.

To understand the settling process better, we need to 
incorporate the radion potential term $f_0(r)$ of the Jordan-frame
potential (\ref{eq:potent}) in out analyses.
If $\vbulk$ is the radion potential in $4+n$ dimensions,
then, after integrating over the $n$ extra dimensions, the
corresponding Jordan frame potential is 
\beq
f_0(r) = r^n\vbulk
\label{f0def}
\eeq
[see Appendix \ref{sec:jordein}], and the Einstein frame potential
$V_E$ obtained after conformal transformation is
\be
\ve=\rzero^{2n}r^{-n}\vbulk={\rzero^n\exp(-\Phi/\mu)
\vbulk}={\rzero^n\vbulk\over\sqrt{u}}.
\ee
Suppose that 
\be
\vbulk=\Ubulk F(r),
\label{f0defa}
\ee
where $\Ubulk$ is a constant and $F(r)$ is dimensionless.
We assume that $F(r)$ may undulate up and down, but with
a characteristic amplitude $\vert F(r)\vert\sim 1$; thus
the scale of the radion potential is determined by 
$\Ubulk$. Since $F(r)$ is dimensionless, it must contain
mass scales; these are reflected in the magnitude(s) of the
derivative(s) of the potential. We assume that $F(r)$ may
have multiple minima (an infinite number in the model
considered below.)

One condition for the radion to be able to settle into
one of the minima of its potential is that the Einstein
frame kinetic energy density fall below the Einstein
frame potential energy amplitude. Since the kinetic energy
is
\be
{\mu^2\over 8}\biggl({u^\prime\over u}\biggr)^2
={\mu^2\over 2\tb^2},
\ee
and the potential energy amplitude is
\be
{\rzero^n\Ubulk\over\sqrt{u}}={\rzero^n\Ubulk\mu\over 2\tb
\sqrt{\Vhat1}},
\ee
during the powerlaw regime, the two become comparable at
at an Einstein frame time
\be
\tb={\mu\sqrt{\Vhat1}\over\rzero^n\Ubulk}.
\ee
This time ought to be smaller than or comparable to the
time at which $u\to 1$ i.e. we must require
\be
{\mu\sqrt{\Vhat1}\over\rzero^n\Ubulk}\sim
{\mu\over 2\sqrt{\Vhat1}},
\ee
which implies
\be
\rzero^n\Ubulk\sim 2\Vhat1.
\label{eq:ubulkrzn}
\ee
Since, presumably, $\Ubulk$ and $\Vhat1$ are determined by fundamental
physics, this relationship may be taken to determine $\rzero^n$.
Moreover, since we know that $\mpl^2\sim m_s^{2+n}\rzero^n$, we 
find
\be
\mpl^2\sim {2\Vhat1 m_s^{2+n}\over\Ubulk}\qquad{\rm or}\qquad
\Ubulk\sim {2\Vhat1 m_s^{2+n}\over\mpl^2}.
\label{eq:ubulkrzn1}
\ee
If we assume that $\Vhat1\sim m_{EW}^4$, where $m_{EW}$ is the
electroweak unification scale, and $m_s>m_{EW}$, then it is
clear that $\Ubulk/m_s^{4+n}\sim{m_{EW}^4/\mpl^2m_s^2}\ll 1$.
In getting these estimates, we have presumed that $F(r)$ takes
on a typical value for $r\sim\rzero$ i.e. we are excluding
the possibility that, for example $\vert F(r\sim\rzero)\vert\ll 1$,
which would alter the above estimates. This amounts to assuming
that whatever the mass scales appear in $F(r)$
are generally of order $\rzero^{-1}$ or larger.

To investigate the settling process in more detail, we need
the equations of motion, which can be obtained from Eqs.\
(\ref{basic:einstein}) and (\ref{basic:radion}).  The resulting
equations are
\baray
\Hbar^2={8\pi\over 3\mpl^2}\biggl[{\rho\over u}+{\Vhat1\over u}
+{\mu^2(u^\prime)^2\over 8u^2}+{\Upsilon\Vhat1\over\sqrt{u}}
F(\rzero u^{1/2n})\biggr]
\nonumber\\
u^{\prime\prime}+\biggl(3\Hbar-{u^\prime\over u}\biggr)
{u^\prime\over u}={4\Vhat1\over\mu^2}
-{2\Upsilon\Vhat1\sqrt{u}\over\mu^2}
\biggl[{r\over n}{dF(r)\over dr}-F(r)\biggr]_{r=\rzero u^{1/2n}}.
\earay
These equations are augmented by the conservation condition 
$\rho\ab^4/u=$constant.  Also we have defined
\be
\Upsilon\equiv{\rzero^n\Ubulk\over\Vhat1},
\ee
which we expect to be $\sim 2$. If we nondimensionalize as before
we find
\baray
\biggl({y^\prime\over y}\biggr)^2={1\over 2\nu}\biggl({\zeta^\prime\over
\zeta}\biggr)^2+{v\over\zeta}+{1\over y^4}
+{v\Upsilon\sqrt{u_i}\over\zeta^{1/2}}F(r_i\zeta^{1/2n})\qquad\qquad
\qquad\nonumber\\
\zeta^{\prime\prime}+\biggl({3y^\prime\over y}-{\zeta^\prime\over
\zeta}\biggr)\zeta^\prime=\nu v-{\nu v\Upsilon\sqrt{u_i}\over 2}
\zeta^{1/2}\biggl[{r\over n}{dF(r)\over dr}-F(r)\biggr]_{r=r_i\zeta^{1/2n}}.
\earay
Notice that in this form of the equations, $\Upsilon$ only appears in
the combination $\Upsi\equiv\Upsilon\sqrt{u_i}$, and since we expect $u_i\ll 1$,
this parameter is small, implying that deviations from the powerlaw
solution only appear at late times, as we have already concluded.

These equations contain three parameters explicitly: $n$ (or $\nu$),
$v=\Vhat1/\rho_i$, and $\Upsi$. In addition, they
contain one (or more) parameters implicitly, because of the mass
scales implicit in $F(r)$. For example, if
\be
F(r)=1-\cos(m_rr),
\label{f0defb}
\ee
so that there is only one mass scale, $m_r$, there is an additional
nondimensional parameter $\mu_i\equiv m_rr_i$. For this form of $F(r)$,
the evolution equations are
\baray
\biggl({y^\prime\over y}\biggr)^2={1\over 2\nu}\biggl({\zeta^\prime\over
\zeta}\biggr)^2+{v\over\zeta}+{1\over y^4}
+{v\Upsi\over\zeta^{1/2}}
[1-\cos(\mu_i\zeta^{1/2n}]\qquad\qquad\qquad\nonumber\\
\zeta^{\prime\prime}+\biggl({3y^\prime\over y}-{\zeta^\prime\over
\zeta}\biggr)\zeta^\prime=\nu v-{\nu v\Upsi\over 2}
\zeta^{1/2}\biggl[{\mu_i\zeta^{1/2n}\sin(\mu_i\zeta^{1/2n})\over n}
-1+\cos(\mu_i\zeta^{1/2n})\biggr].
\label{eq:cosmodel}
\earay
In these units, substantial deviations from the scaling equations
are expected after 
\be
\taub=\Upsi^{-1}\sqrt{n+2\over 12nv},
\ee
at which time
$\zeta=4/\Upsi^2$, provided that 
$\mu_i\gta(\Upsi/2)^{1/n}$.
 
Fig. 2 shows the results of numerically integrating the dimensionless
evolution equations, eqs. (\ref{eq:cosmodel}),
for $(v,\Upsi,\mu_i)=(10^{-6},10^{-8},10^{-3})$ (Fig. 2a) 
and $(v,\Upsi,\mu_i)=(10^{-4},10^{-6},10^{-2})$ 
(Fig. 2b), respectively, with $n=2$ in both
cases. The numerical results show clearly that after a long period
of powerlaw expansion (in close agreement with the fixed point
solution found above), $\zeta$ levels off,
although in both cases, the time at which this happens is a bit
later than our back-of-envelope estimate, so that $\zeta$ is
systematically larger than $4/\Upsi^2$ asymptotically. For
$\Upsi=10^{-6}$, we would estimate $\zeta=4\times 10^{12}$
asymptotically, whereas the numerical result is $9.75\times
10^{13}$, a factor of about 25 larger; for $\Upsi=10^{-8}$
we would estimate $\zeta=4\times 10^{16}$, as opposed to the
numerical $4\times 10^{17}$, about a factor of 10 larger.
(The discrepancy in $r$ is smaller, since
$\zeta\propto r^{2n}=r^4$ for $n=2$.)
These results can be explained if the time to asymptote is
a factor of three to five larger than our simple estimate.
By inspecting the figures, we can see that the time at
which $\zeta$ levels off is about five times larger than
the analytic estimate, $\approx 4.08\times 10^7$, for
$(v,\Upsi,\mu_i)=(10^{-4},10^{-6},10^{-2})$, and about
three times larger than the analytic estimate,
$\approx 4.08\times 10^{10}$, for 
$(v,\Upsi,\mu_i)=(10^{-6},10^{-8},10^{-3})$.

Once the radion field begins to settle into a minimum
around $\rzero$, the value of the Planck mass zeros in
on its present value. \footnote{It turns out that for 
a quadratic potential, there is also an exact solution
in the radiation dominated era if we ignore the source
term. In this case, the equations are completely linear,
and if we take a potential like ${1\over 2}m^2
(\Phi-\Phi_0)^2$ the solution is
\be
\Phi-\Phi_0={A_+J_{1/4}(m\tb)+A_-J_{-1/4}(m\tb)
\over (m\tb)^{1/4}};
\ee
the most important feature of this equation is that the
amplitude falls like $(m\tb)^{-3/4}$ (i.e. $\propto
\ab^{-3/2}$) at late times.}
Once this happens, the temperature
of the Universe can begin to fall once more, and 
ultimately it must drop below $\Tc1$. When this happens,
the inflaton fields once again are free to roll, and
move toward their minimum at nonzero vacuum expectation
values. The amplitude of any residual oscillations in
the radion field will then redshift away exponentially,
until the inflaton kinetic energy is thermalized in a
second phase of reheating. It is therefore necessary
that once the temperature of the Universe becomes
constant during the radiation-dominated era of radion
growth, the constant temperature must be above $\Tc1$.
Moreover, we need to require that there is enough inflation
after the temperature falls below $\Tc1$ for the amplitude
of oscillations in radius to drop to an acceptable level.

%%%%%%%%%%%%%%%%%%%%%%%%%%%%%%%%%%%%%%%%%%%%%%%
%\begin{figure}[t]
\begin{figure}[b]
\begin{center}
\hspace{1cm}
%\vspace*{}
\epsfxsize=11 cm
%\epsfysize=9 cm
%\epsfbox{trial2.eps}
\epsfbox{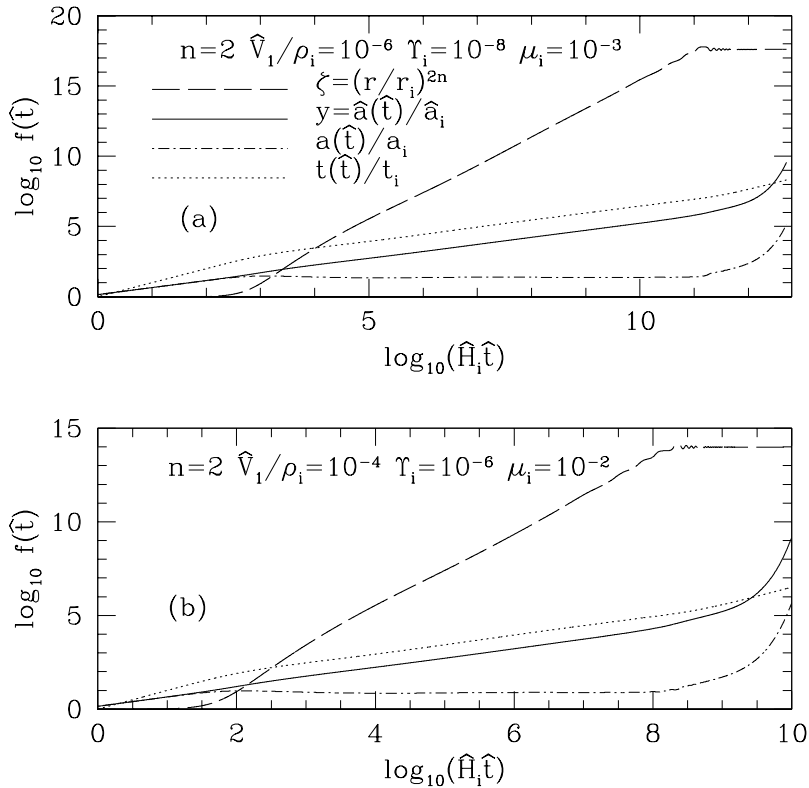}
%\bigskip
\caption{
Numerical solutions for the cosmological model
during Phase II for two different sets of parameters.
The long-dashed line is $\zeta(\hat t)$,
the solid line is $\hat a(\hat t)$,
the dot-dashed line is
$a(\hat t)$, and the dotted
line is $t(\hat t)$.
Both solutions are characterized by an initial adjustment
period during which the radiation energy density, which
is proportional to $a_J^{-4}$, drops,
followed by a protracted period of powerlaw growth of
the radius of the extra dimensions (at fixed temperature),
which terminates when the radion potential traps the field
at a minimum. Exponential inflation
begins once the radion becomes trapped.
}
\end{center}
\label{fig:scenario}
\end{figure}
%%%%%%%%%%%%%%%%%%%%%%%%%%%%%%%%%%%%%%%%%%%

For $F(r)=1-\cos m_rr$, it is easy to see that the 
minima of $V_E(\Phi)$ are at $m_rr=2\pi k_r$, where
$k_r$ is an integer. At minima,
\be
{d^2V_E(\Phi)\over d\Phi^2}={4\pi^2k_r^2
\rzero^n\Ubulk\over\mu^2 n^2}
\biggl({\rzero\over r}\biggr)^n,
\ee
and, at the minimum corresponding to $r=\rzero$,
\be
{d^2V_E(\Phi)\over d\Phi^2}={4\pi^2k_{\rzero}^2
\rzero^n\Ubulk\over\mu^2 n^2}.
\ee
Thus, near $r=\rzero$ (or $\Phi=0$)
\be
V_E(\Phi)\approx{2\pi^2k_{\rzero}^2
\rzero^n\Ubulk\over\mu^2 n^2}\Phi^2,
\ee
and the mass of the radion is
\be
m^2_{\rm radion}=
{4\pi^2k_{\rzero}^2
\rzero^n\Ubulk\over\mu^2 n^2}
\sim{8\pi^2 k_{\rzero}^2\Vhat1\over\mu^2 n^2}
={256\pi^3 k_{\rzero}^2\Vhat1\over n(n+2)
\mpl^2};
\ee
numerically, we find (recall Eq.\ [\ref{eq:ubulkrzn}])
\be
m_{\rm radion}\sim 7.3\times 10^{-3}\,{\rm eV}
{\,k_{\rzero}\over\sqrt{n(n+2)}}
\biggl({\Vhat1\over 1\,{\rm TeV^4}}\biggr)^{1/2}.
\label{eq:radionmass}
\ee
Remarkably, the mass of the radion that emerges is
much smaller than any other characteristic mass
scale in the problem, unless $k_{\rzero}\gg 1$
and/or $\Vhat1\gg 1\,{\rm TeV^4}$.
For other choices of $F(r)$ we would have
\be
{d^2V_E(\Phi)\over d\Phi^2}={\rzero^n\Ubulk\over
\mu^2n^2}\biggl({\rzero\over r}\biggr)^n
\biggl[-n(n-1)F(r)+r^2{d^2F(r)\over dr^2}\biggr];
\ee
with the definition
\be
(2\pi k_r)^2\equiv
\biggl[-n(n-1)F(r)+r^2{d^2F(r)\over dr^2}\biggr]
\ee
this becomes identical to the formula for the
special case $F(r)=1-\cos m_rr$.

\section{Expansion Factors and The Radion Density}
\label{sec:exprad}

Now that we have a complete account of the various phases of
expansion in our proposed cosmological model, we can gather
the results to calculate the factors by which the Universe
has expanded between various interesting epochs and the
present. We shall work in the Jordan frame, for the most
part, and define the cosmological scale factor $a(t_0)=1$
at the present day, $t=t_0$. The value of the Hubble constant
today is $H_0=100h_0\,{\rm km\,s^{-1}\,Mpc^{-1}}$, 
the CMBR temperature today is $T_0 = 2.7^\circ \, {\rm K}$, and the
corresponding critical density is $\rco=3H_0^2\mpl^2/8\pi
\approx 8.01\times 10^{-47}h_0^2\Gev^4$.

Before moving on to our more complicated cosmological model,
it is useful to review the situation in convention cosmology
with a fixed Planck mass and a single inflationary era. Let
$\aexp$ be the value of the scale factor at the end of the
period of exponential expansion, and $\arh$ the value of the
scale factor at the end of the reheating phase that follows
exponential expansion. If $\trh$ is the reheating temperature,
then
\be
\arh={\so T_0\over\srh\trh},
\ee
where the dimensionless factors $\scal_0$ and $\scal_{rh}$ count particle
states in thermodynamic equilibrium at present and at $\arh$, respectively.
\footnote{Recall that it is entropy that is conserved during adiabatic
expansion. Note that $\scal_{rh}$ may be considerable e.g. $\gta 10$.} 
If the energy density during inflation is 
$\rho_V$, and the inflaton potential is harmonic near the minimum
attained at the end of inflation, then
\be
{\pi^2\erh\trh^4\over 15}\sim\rho_V\biggl({\aexp\over\arh}
\biggr)^3,
\ee
where $\erh$ is another dimensionless factor that counts the contributions
of various particle states to the total energy density at temperature
$\trh$. \footnote{When only relativistic particles are present, and
the equation of state depends only on temperature, $\erh=3\scal_{rh}/4$.}
Defining
\be
\epsrh\equiv\biggl({\pi^2\erh\trh^4\over 15\rho_V}\biggr)^{1/4},
\ee
we find $\aexp/\arh\sim\epsrh^{4/3}$, or
\be
\aexp\sim{\epsrh^{4/3}\so T_0\over\srh\trh}
={\epsrh^{1/3}\so\erh^{1/4}\over\srh}
\biggl({\pi^2T_0^4\over 15\rho_V}\biggr)^{1/4}.
\ee
If we assume that the present day Hubble scale $H_0^{-1}$ passed outside
the horizon during inflation, then the scale factor of the Universe at
that time was $\ahub=(\rco/\rho_V)^{1/2}$, and the ratio
\be
{\aexp\over\ahub}\sim{\epsrh^{1/3}\so\erh^{1/4}\over\srh}
\biggl({\pi^2T_0^4\rho_V\over 15\rco^2}\biggr)^{1/4}
=\exp\biggl[30.8+0.25\ln\biggl({\rho_V\over 1\,\Tev^4}\biggr)
+\ln\biggl({\epsrh^{1/3}\so\erh^{1/4}\over\srh h_0}\biggr)
\biggr].
\label{eq:expfacconv}
\ee
For $\rho_V\approx (10^{15}\Gev)^4$, Eq.\ (\ref{eq:expfacconv})
implies about 60 e-foldings between $\ahub$ and $\aexp$, the
familiar value, but generally the number of e-foldings depends on
details of the inflationary model.

In the cosmological model developed above, there are two periods
of exponential inflation that occur at different values of the
Planck mass. The second inflationary epoch, and the subsequent
reheating, occur at the end of Phase II, at which time the Planck
mass has settled to its present value. Thus, we can apply the same
reasoning to this epoch as was developed in the preceding paragraph
and we find that the scale factor at the end of the period of 
inflation that concludes Phase II is
\be
\aexptwo\sim{\epsrhtwo^{1/3}\so\erhtwo^{1/4}\over\srhtwo}
\biggl({\pi T_0^4\over 15\Vhat1}\biggr)^{1/4}
\ee
in the Jordan frame,
where $\epsrhtwo$, $\erhtwo$ and $\srhtwo$ have the same meanings
as the analogous symbols introduced for conventional inflation
and reheating, but apply to the end of Phase II only. If we assume
that this second inflation led to an increase in scale by a factor
$\fcal_2$, then the scale factor just before inflation began was
\be
a_2={\aexptwo\over\fcal_2}\sim{\epsrhtwo^{1/3}\so\erhtwo^{1/4}\over\srhtwo\fcal_2}
\biggl({\pi T_0^4\over 15\Vhat1}\biggr)^{1/4};
\ee
this is also the value of the scale factor before the radion field
began its powerlaw growth during Phase II. 

Proceeding backward in time still further, we encounter the first subphase
of Phase II, during which the energy density of the Universe declined from
its value just after the reheating at the end of Phase I, 
$\pi^2\erhone\trhone^4/15$, to $\sim\Vhat1$; the corresponding increase
in scale was a factor $\sim(\pi^2\erhone\trhone^4/15\Vhat1)^{1/4}$, so
the scale factor at the end of the reheating that terminated Phase I
was
\be
\arhone\sim{\epsrhtwo^{1/3}\so\erhtwo^{1/4}T_0
\over\srhtwo\fcal_2\erhone^{1/4}\trhone}
={\epsrhtwo^{1/3}\so\erhtwo^{1/4}\over\srhtwo\fcal_2\epsrhone}
\biggl({\pi^2 T_0^4\over 15\Vhatz}\biggr)^{1/4},
\ee
where $\epsrhone\equiv(\pi^2\erhone\trhone^4/15\Vhatz)^{1/4}$.
Consequently, the value of the cosmological scale factor at the
end of the exponential expansion in Phase I is
\be
\aexpone\sim\epsrhone^{4/3}\arhone
\sim{\epsrhone^{1/3}\epsrhtwo^{1/3}\so\erhtwo^{1/4}\over\srhtwo\fcal_2}
\biggl({\pi^2 T_0^4\over 15\Vhatz}\biggr)^{1/4}.
\ee
Although the derivation of $\aexpone$ is more complicated than
the derivation of $\aexptwo$ (or its conventional equivalent,
$\aexp$), notice that it does not depend explicitly on the different
Planck scales that arise in our inflationary model. 

The dependence on Planck scales enters when we reconsider the
relationship between $\aexpone$ and $\ahub$. We have assumed
that the present Hubble scale, and all other macroscopic scales
relevant to the development of large scale structure, crossed
the horizon during Phase I. As a result, 
$\ahub=(\rco/\Vhatz)^{1/2}(\mplI/\mpl)$, and
\baray
{\aexpone\over\ahub}\sim
{\epsrhone^{1/3}\epsrhtwo^{1/3}\so\erhtwo^{1/4}\mpl\over\srhtwo\fcal_2\mplI}
\biggl({\pi^2 T_0^4\Vhatz\over 15\rco^2}\biggr)^{1/4}
\qquad\qquad\qquad\qquad\qquad\qquad\qquad\qquad
\nonumber\\
\qquad\qquad
=\exp\biggl[30.8+0.25\ln\biggl({\Vhatz\over 1\,\Tev^4}\biggr)
+\ln\biggl({\mpl\over\fcal_2\mplI}\biggr)
+\ln\biggl({\epsrhone^{1/3}\epsrhtwo^{1/3}\so\erhtwo^{1/4}\over\srhtwo
h_0}\biggr)\biggr].
\label{eq:efolds}
\earay
For our cosmology, the number of e-foldings between horizon crossing and the
end of exponential expansion during Phase I depends on numerous uncertain
parameters, principally $\Vhatz$ and the combination
$\mpl/\fcal_2\mplI$. The requirement that $\aexpone>\ahub$ is one
constraint on our cosmological model.

We can obtain a separate constraint 
by requiring that the energy density in radions today
does not overfill the Universe. 
At the end of the subphase of powerlaw growth of
the radion during Phase II, the energy density in radions is $\sim\Vhat1$;
during the ensuing exponential expansion it drops to $\sim\Vhat1/\fcal_2^3$.
The energy density in radions drops by an additional factor of
$(\aexptwo/\arhtwo)^3\sim\epsrhtwo^{4/3}$ by the time reheating is complete, 
to a value
$\sim\pi^2\erhtwo\trhtwo^4/15\fcal_2^3$ i.e. a factor $\sim\fcal_2^3$
smaller than the energy density in relativistic matter at the end of
reheating. Between $\trhtwo$ and $T_0$, the radion density drops by a
factor of $\scal_0T_0^3/\scal_{rh,2}\trhtwo^3$, so that
\be
\rho_{rad,0}\sim{\pi^2\scal_0\erhtwo\trhtwo T_0^3\over 15\scal_{rh,2}
\fcal_2^3}
\ee
is the density of radions today. Comparing with the closure density
implies a radion density parameter
\be
\Omega_{rad,0}\sim{8\pi^3\scal_0\erhtwo\trhtwo T_0^3\over
15\scal_{rh,2}\fcal_2^3H_0^2\mpl^2}
\approx {(\trhtwo/\fcal_2^3)(\scal_0\erhtwo/\scal_{rh,2}h_0^2)\over
10\,{\rm eV}};
\label{eq:omegaradion}
\ee
thus, $\Omega_{rad,0}\lta 1$ as long as $\trhtwo/\fcal_2^3\lta
10(h_0^2\scal_{rh,2}/\scal_0\erhtwo)\,{\rm eV}$.

\section{Fluctuations}
\label{sec:fluct}

Fluctuations about the smooth cosmological background alter
the form of the Jordan frame line element from Eq.\ (\ref{eq:jbkgnd})
to
\be
ds^2=-dt^2+a^2(t)(\delta_{ij}+h_{ij})dx^idx^j;
\ee
the corresponding Einstein frame metric is 
$d\sb^2=ds^2\sqrt{u(\xvec,t)}$, where
$u(\xvec,t)\equiv\exp[2\Phi(\xvec,t)/\mu]$.
Assume that $u(\xvec,t)=u_0(t)[1+\eta(\xvec,t)]$,
where $\vert\eta(\xvec,t)\vert\ll 1$. Then after making an
appropriate infinitesimal coordinate transformation, we
get
\be
d\sb^2=-d\tb^2+\ab^2(\tb)(\delta_{ij}+\bh_{ij})
d\xb^id\xb^j
\label{eq:metpert}
\ee
i.e. the metric in the Einstein frame can be reduced to 
synchronous form. As always, the necessary infinitesimal
coordinate transformation is not unique; there is
still gauge freedom even when the metric is reduced to 
the form of Eq.\ (\ref{eq:metpert}) \cite{swein,gauge}.

The perturbed Ricci tensor corresponding to Eq.\ (\ref{eq:metpert})
can be found in Ref. \cite{swein}. For compressional perturbations
around the powerlaw background solution for Phase II,
the perturbation equations can be boiled down to two gauge-independent
equations,
\baray
\phihbdp+{5\phihbp\over 2\tb}-{\nablab^2\phihb\over 3\ab^2}
=-\biggl({n+2\over 24n}\biggr){\psi\over\tb}\nonumber\\
{\nablab^2\phihb\over\ab^2}=-\biggl({9n-6\over 32n}\biggr)
{(\Delta-\chi)\over\tb^2}-\biggl({n+2\over 16n}\biggr){\psi\over
\tb},
\label{eq:gaugepot}
\earay
where $\phihb$ is the (Einstein-frame)
gauge-independent metric potential introduced
in Ref. \cite{gauge} and $\psi=\eta^\prime+\eta/\tb$,
coupled to three gauge-dependent equations,
\baray
\etadp+{3\etap\over 2\tb}+{\eta\over\tb^2}-{\nablab^2\eta\over\ab^2}
=-{\bhp\over t}\nonumber\\
\bhdp+{\bhp\over t}=-\biggl[\biggl({9n-6\over 8n}\biggr){\Delta\over
\tb^2}+\biggl({n+2\over 8n}\biggr){\eta\over\tb^2}
+\biggl({n+2\over 2}\biggr){\etap\over\tb^2}\biggr]\nonumber\\
(\chi t)^\prime+{\Delta-\chi\over 2}=0,
\label{eq:gaugepert}
\earay
where $\bh=\bh_{ii}$.
If $\rho_0(\tb)$ and $u_0(\tb)$ denote the background
solutions given by eqs. (\ref{eq:rhofix}) and (\ref{eq:ufix}),
\be
\rhobar={\rho(\xbvec,\tb)\over u(\xbvec,\tb)}
={\rho_0(\tb)\over u_0(\tb)}[1+\Delta(\xbvec,\tb)],
\ee
and, if the Einstein-frame three-velocity is $\Uvec=\gradb V(\xbvec,\tb)$,
\be
\chi(\xbvec,\tb)={2\ab^2(\tb)V(\xbvec,\tb)\over\tb}.
\ee
Eqs. (\ref{eq:gaugepert}) and the second of eqs. (\ref{eq:gaugepot})
can be combined to yield the gauge-independent
equation
\be
\psitp+{13\psidp\over 2\tb}+\biggl({37n-2\over 4n}\biggr){\psip\over\tb}
+\biggl({9n-2\over 4n}\biggr){\psi\over\tb^3}
-{3\nablab^2\psi\over\tb\ab^2}-{\nablab^2\psip\over\ab^2}=
-{6\nablab^2\phihb\over\tb^2\ab^2}-{4\nablab^2\phihbp\over\tb\ab^2},
\label{eq:psieq}.
\ee
On scales larger than the Einstein-frame horizon scale, $\Hbar^{-1}$,
a complete solution may be obtained by coupling Eq.\ (\ref{eq:psieq})
to the first of eqs. (\ref{eq:gaugepot}).

At the end of Phase I, there are no fluctuations in the radion field,
so $\psi=0$ on all scales of interest today (i.e. well outside the
horizon). (This is because the effective mass of the radion field
once in a minimum is generally larger than the cosmological expansion
rate; for a particular example, see Appendix \ref{sec:phase0more}.)
There are fluctuations in $\phihb$ on these scales, and,
from Eq.\ (\ref{eq:psieq}), these tend to generate perturbations in
the radion field. However, the driving terms are very small on large
scales: for comoving wavenumber $k$, they are of order $k^2\phihb/\tb^2\ab^2
\ll\phihb/\tb^4$. Consequently, $\psi\sim k^2\tb\phihb/\ab^2\ll
\phihb/t$, and the large-scale fluctuations in the radion field generated during
Phase II are far smaller than $\phihb$ (which sets the scale of the
density fluctuations on such scales after they re-enter the horizon).
Moreover, the source term in the equation for $\phihb$ is negligible on these
scales, since $\psi/\tb\ll\phihb/\tb^2$, and $\phihb$ remains constant on
scales larger than the horizon during Phase II.

\section{Discussion}
\label{sec:discussion}

In the preceding section, we developed a new picture for cosmology
in the brane world. Our cosmological model is based on a specific form
of the effective potential, Eq.\ (\ref{eq:potent}), which, although 
admittedly somewhat complicated, allows the size of the compact
dimensions of the Universe to evolve to its present value from
a different, but much smaller, fixed value at early times.
In the specific scenario we have unfolded, the resulting evolution
of the Universe divides naturally into four different phases, the
last of which can be called the ``standard Big Bang cosmology'' that
follows the electroweak phase transition and proceeds to the present
day, with the radius of the extra dimensions fixed at its present
value, $r_0$, and hence the Planck mass fixed at $G^{-1/2}=
1.22\times 10^{19}\Gev$.

The other three phases represent the evolution of the radius of the
extra dimensions to that value from a considerably smaller one. While
we do not claim that the scenario we have developed for this evolution
is unique, it does have some features that are attractive. The first
phase, Phase 0 of \S\ref{sec:phase0}, is relatively brief; the
radion settles into a potential minimum at $\rI$ during this phase.
We have shown that this process is not entirely guaranteed to take
place, but may be rather likely in scenarios where the effective
potential for the radion has multiple (or an infinite number of)
minima. 

Phase 0 sets the stage for Phase I of \S\ref{sec:phase1},
during which the Universe inflates
at a fixed Planck mass $\mplI=\mpl(\rI/r_0)^{n/2}<\mpl$. We assume
that this is the main inflationary phase undergone by the expanding
Universe, so that macroscopic comoving scales on which large
scale structure develops all passed outside the horizon
during Phase 0. The density perturbation amplitude produced by
quantum fluctuations in the inflaton field(s) $\psivec$ during
Phase 0 is estimated in Eq.\ (\ref{eq:densitypert}). For an 
inflaton effective potential proportional to 
$m_s^4[1-\exp(-m_dd)]$, where $d$ is the inter-brane
separation (as discussed in \cite{witten} and Appendix B),
we estimate that the primordial density fluctuation amplitude is
$\sim(m_d/\mplI)N_k$ for a mode with comoving
wavenumber $k$, with $N_k$ the number of e-foldings between
horizon crossing for that mode and the end of exponential
inflation during Phase 0. (In Appendix B, we argue
for $m_d=m_{RR}$, the mass of the RR mode.) Nominally,
we would expect $m_d\lta m_s$, leading to a density perturbation
amplitude $\lta m_s/\mplI$, which would be woefully small
for $m_s\sim\Tev$ if $\mplI=\mpl$. An attractive feature of
our scenario is that it allows $\mplI\ll\mpl$. Turning the
argument around, observations of large scale temperature
fluctuations in the cosmic microwave background radiation
\cite{cmbrefs}
require $m_d/\mplI\sim 10^{-5}$, so the radius of the
extra dimensions during Phase I satisfies the constraint
\be
\biggl({\rI\over r_0}\biggr)^{n/2}
={\mplI\over\mpl}\sim 10^5{m_d\over\mpl}
\approx 10^{-11}\biggl({m_d\over 1\,\Tev}\biggr).
\label{eq:rlim}
\ee
The radius of the extra dimensions during Phase I was considerably
smaller than today if $m_d\sim 1\,\Tev$ and $n\leq 7$. One of
the principal motivating factors behind our cosmological model
is the realization that the amplitude of primordial density
perturbations is proportional to $\mplI^{-1}$, and that perturbations
at an acceptable amplitude are only possible if $\mplI\ll\mpl$.

Expansion from $r_I$ to $r_0$ occurred during Phase II, which
is mainly radiation-dominated following the reheating that 
terminated Phase I. In our model, once the branes come to overlap,
$V_0(\psivec)\to 0$, which frees the radion to expand once more since
the product term $V_0(\psivec) f_I(r)$ in the effective potential
(\ref{eq:potent}) is no longer active.  The inflaton potential is then 
dominated by $V_1(\psivec)$, 
and we assumed that the initial reheating was sufficient to trap
the Universe at small $\psivec$ at first, at a minimum with
nonzero $V_1(\psivec)$. Phase II naturally divides into
three sub-phases, which was discussed in detail in \S\ref
{sec:phase2}. During the first subphase, the Universe expands
at $r\approx r_I$, until it cools sufficiently that an
approximate equilibrium is attained, with comparable
energies in radiation, radion kinetic energy, and vacuum
energy density. Once this happens, a new phase of powerlaw
expansion of the radius of the extra dimensions ensues at
virtually fixed radiation temperature; see eqs. (\ref{eq:ufix})
and (\ref{eq:rhofix}). 
This subphase ends
when the radion becomes trapped in one of the many (or
infinite) minima of its effective potential, $f_0(r)$
(see Eq.\ [\ref{eq:potent}]); we assume that this minimum
is at $r_0$, and eqs. (\ref{eq:ubulkrzn}) and (\ref{eq:ubulkrzn1})
estimate the radion vacuum energy density in the bulk 
required for this to be true. The third subphase of Phase II is
the phase transition associated with the inflaton potential
$V_1(\psivec)$ in Eq.\ (\ref{eq:potent}). During this phase,
the Universe expands exponentially by an additional factor
$\fcal_2$, and reheats, finally, to a temperature
$\trhtwo$. Conventional, non-inflationary Big Bang
cosmology commences at this point.

Another important constraint on our cosmological model is
the requirement that the present day radion energy density
does not dominate the total energy density of the Universe.
Because the radion is trapped in a potential minimum toward
the end of Phase II, it behaves, much like the axion, as
a massive, cold dark matter particle; the effective
radion mass is estimated in Eq.\ (\ref{eq:radionmass}),
and may be $\sim 1\,{\rm eV}$ typically.\footnote{We shall discuss the
development of density perturbations during the ``matter-dominated''
phase of a Universe consisting of radions and other cold dark
matter elsewhere, but note here that there is nothing special
about the radion component, and it can be shown to behave
as a typical dark matter particle. As shown in \S\ref{sec:fluct},
significant, additional fluctuations in the radion field are
not produced during Phase II.} 
Just after the reheating that ends the short inflationary
period during Phase II, the energy density in radions is smaller
than the energy density of the products of reheating by a factor
$\approx\fcal_2^{-3}$. Requiring that radions not dominate
the mass density of the Universe today implies, by
Eq.\ (\ref{eq:omegaradion}),
\be
\fcal_2\gta 2\times 10^3\biggl({\trhtwo\over 100\,\Gev}\biggr)^{1/3}
\biggl({\scal_0\erhtwo\over h_0^2\srhtwo}\biggr)^{1/3},
\label{eq:f2lim}
\ee
where $\trhtwo$ is the temperature of the Universe after this
last reheating episode, and the remaining factors are
$\sim 1$ in general; see \S\ref{sec:exprad} for details.
\footnote{This also guarantees that the radion energy density
during cosmological nucleosynthesis was no more important than
that of any other dark matter component, and therefore has
negligible effect.} Thus,
if the second inflationary epoch comprised more than about 
eight e-foldings, the present day density in radions would
be negligible, but our model cannot be consistent with fewer
than eight e-foldings, which would result in an overdense
Universe dominated by radions. We note that Newton's constant
of gravitation actually oscillates in this model at a frequency
$m_{\rm radion}$, but with a very small amplitude,
$\delta G/G\sim \Omega_{rad,0}^{1/2}H_0/m_{\rm radion}$
(e.g. \cite{swill}).

Combining eqs. (\ref{eq:efolds}), (\ref{eq:rlim}) and
(\ref{eq:f2lim}), we can constrain the number of e-foldings
that take place during Phase I between the time when the
present-day Hubble length passed outside the horizon and
the end of exponential expansion:
\baray
{\aexpone\over\ahub}\sim
\exp\biggl[48.7+0.25\ln\biggl({\Vhatz\over 1\,\Tev^4}\biggr)
+\ln\biggl({1\Tev\over m_d}\biggr)\biggl({100\,\Gev\over\trhtwo}\biggr)^{1/3}
+\ln\biggl({\epsrhone^{1/3}\epsrhtwo^{1/3}\Omega_{rad,0}^{1/3}\over
h_0^{1/3}\epsrhtwo^{1/12}}\biggr)
\biggr].
\label{eq:expconstrain}
\earay
While we expect this to be below the 60 e-foldings generally found
for GUT-scale inflation, it need not be far smaller (e.g. by a factor of
two), as one might have expected for a theory in which inflation happens
at a much lower energy scale (e.g. $\sim 1\,\Tev$ compared to 
$\sim 10^{12}\Tev$). This is because the Planck scale was relatively
small during Phase I, when the main inflationary era occurred in
our model. 
According to Eq.\ (\ref{eq:expconstrain}), it is not
too difficult to satisfy the constraint that the number of e-foldings
be considerably larger than one,
unless $\Omega_{rad,0}$ is absurdly small.

Ours is only one of several proposed scenarios for cosmology
in the brane world. The model expounded here
has some overlap with that of Ref \cite{adkm}, except that
we assume that the radion and inflaton are different fields,
resulting in substantial differences between the two models.
By contrast to our model, Ref\cite{az} uses a bulk potential for the
inflaton, leading to substantially different physics, and
Ref\cite{bend} proposes a different theory of baryogenesis whereas we
believe that baryogenesis from the (minimal  
supersymmetric) standard model electroweak phase transition is adequate.
It should be clear to the readers that there may exist other viable 
cosmological scenarios in the brane world. Eventually, string theory 
should provide the appropriate radion/inflaton potentials, which hopefully 
will determine the cosmological scenario that nature chooses.

Recently, Randall and Sundrum proposed a scenario \cite{rs} 
where the extra dimension does not have to be compactified. In this 
scenario, the radius can have a run-away behavior. It is not even 
clear that the radion field has to be trapped during inflation to 
obtain the correct power spectrum of the density perturbation. 
The cosmology of such a scenario will be interesting to study.
Some attempts along this direction can be found in Ref.\ \cite{rscos}.

\acknowledgments

{}We thank Philip Argyres, Keith Dienes and Lawrence Kidder
for discussions. We thank Nima Arkani-Hamed, Savas Dimopoulos, 
Gia Dvali, Nemanja Kaloper and John March-Russell for informing us 
that they have a similar (but different) scenario. 
\'E.F.\ was supported in part by NSF grant PHY 9722189 and by the
Alfred P. Sloan foundation.  The research of S.-H.H.T. is partially
supported by the National Science Foundation. I.W. gratefully
acknowledges support from NASA.

\appendix

\section{The Effective Potential}
\label{sec:effpot}

Here we want to give some background on the motivation on the 
choice of the effective potential (\ref{eq:potent}) used in the text.
Although our universe is non-supersymmetric, it is very helpful to start 
from a supersymmetric theory in which spontaneous supersymmetry breaking
takes place dynamically.
Since the brane world picture is naturally realized in string theory, 
where many non-trivial consistency properties (such as consistent 
quantum gravity) are automatically built in, we shall consider 
what stringy properties tell us about the effective potential. 

Since string theory has no free parameter (the string scale $m_s$ simply 
sets the mass scale), all physical parameters emerge as various scalar 
fields obtaining vacuum expectation values ($vev$) determined by
string dynamics. For example, the large radii of the large extra dimensions
come from the $vev$s of the radion fields.
Before supersymmetry breaking and dilaton stabilization (the latter fixes 
the string coupling value), the dilaton and the compactification radii are 
moduli, that is, the effective potential is flat (and remains zero) as their 
$vev$s vary.
This is true to all orders in the perturbation expansion. So 
supersymmetry breaking and moduli stabilizations are expected to come 
from non-perturbative dynamics, which is poorly understood at the moment.
However, it is still reasonable to assume that the moduli degeneracy is 
lifted after dynamical supersymmetry breaking.

Although the effective potential of a particular string vacuum 
({\em i.e.}, ground state) is model-dependent, there are stringy and 
supersymmetric features that are quite generic \cite{ibanez}. 
Here we shall give a very brief description of 
some of the properties that are relevant in this paper.
Besides the graviton, the dilaton and the radii, a typical semi-realistic 
string model has gauge fields (in vector super-multiplets), charged 
matter fields (as components of chiral super-multiplets) as well as 
additional moduli. The general Lagrangian coupling $N=1$ supergravity to
gauge multiplets and chiral multiplets $z_i$ (the index $i$ labeling 
different chiral multiplets $z_i$ will be suppressed)
depends on three functions : \\
(1) The K\"ahler potential $K (z,\bar{z})$ which is a {\it real}
function.  It determines the kinetic terms of the chiral fields
\begin{equation}
{\cal{L}}_{kin} = K_{z \bar{z}} \partial_\mu z \partial^{\mu} \bar{z}
\end{equation}
with $K_{z \bar{z}} \equiv \partial^2 K/\partial z \partial
\bar{z}$.  

(2) The superpotential $W (z)$ is a {\it holomorphic} function
of the chiral multiplets (it does not depend on $\bar{z}$).  $W (z)$
determines the Yukawa couplings as well as the $F$-term part of the
scalar potential $V_F$ :
\begin{equation}
V_F (z,\bar{z}) = e^{K/M^2_{Pl}} \left\{ D_z W K^{-1}_{z \bar{z}}
{\overline{D_zW}} - 3 \frac{\mid W \mid^2}{M^2_{Pl}}      \right\}~~,
\end{equation}
with $D_z W \equiv \partial W/\partial z + W K_z/M^2_{Pl}$. 

(3) The gauge kinetic function $f_{ab}(z)$ is also {\it
holomorphic}. It determines the gauge kinetic terms
\begin{equation}
{\cal{L}}_{gauge} = {\rm Re} f_{ab} F^a_{\mu \nu} F^{\mu \nu b} + {\rm
Im} f_{ab} F^a_{\mu \nu} {\tilde{F}}^{\mu \nu b}
\end{equation}
It also contributes to gaugino masses and the gauge part of
the scalar potential $V_D$ :
\begin{equation}
V_D  =  \left( {\rm{Re}} f^{-1} \right)_{ab} (K_z, T^a z) \left(
K_{\bar{z}}, T^b \bar{z} \right) 
\end{equation}
So the effective potential is given by
\begin{equation}
V  =  V_F + V_D
\end{equation}

Consider a semi-realistic Type I string model, {\em i.e.},
a $D=4$, ${\cal N}=1$ supersymmetric,
chiral model, with a set of $9$-branes and up to 3 sets of $5$-branes, 
with a common 4-dimensional uncompactified spacetime ($x_0$ to $x_3$). 
We shall treat the 6 compactified dimensions as composed of 3 
(orbifolded) two-tori: the first torus with coordinates
($x_8$,$x_9$), the second with coordinates ($x_6$,$x_7$)
and the third with coordinates ($x_4$,$x_5$), the volumes of which are, 
crudely speaking, $r_1^2$, $r_2^2$ and $r_3^2$ respectively. 
The 4-dimensional Planck mass $M_{Pl}$ and the
Newton's constant $G_N$ are given by
\begin{equation}\label{newton}
G_N^{-1} = M_{Pl}^2 \sim
{{ (m_s^4 r_1 r_2 r_3)^2} \over \lambda^2}
\end{equation}
where $\lambda$ is the string coupling.
The gauge couplings $g_9$ and $g_{5i}$ of the gauge groups 
$G_9$ and $G_{5i}$ are
\begin{equation}
g_9^{-2} = {{2(m_s^3 r_1 r_2 r_3)^2} \over \lambda} ~, \quad \quad
g_{5i}^{-2} = {2{m_s^2 r_i^2} \over \lambda}
\end{equation}
where the $i$th set of $5$-branes has $r_i$ as the size of its two 
compactified directions.
For large radius $r_3$, $g_9$ and $g_{53}$ are too small to be relevant, so 
the standard model gauge groups must come from the first two sets of 
$5$-branes. This is the $n=$2 case. In the two examples \cite{bw,kt} that 
we know, this $n=$2 case is needed for phenomenology. 
Here, we shall identify $r=r_3$ as the radion.
To stabilize the moduli $vev$s (and maybe also to induce SUSY breaking), the
string coupling $\lambda$ is likely to be strong.
To obtain the weak standard model gauge couplings from
a generic strong string coupling requires that
$m_s r_1$ and maybe $m_s r_2$ to be around 10. This will modify 
Eq.\ (\ref{planckscale}). 
(In semi-realistic string models, the
picture is somewhat more complicated.)

The dilaton and the volume moduli are bulk modes :
\begin{equation}
 S = g_9^{-2} + i \theta , \quad \quad
 T_i = g_{5i}^{-2} + i \theta _i
\end{equation}
where the $\theta$'s are corresponding axionic fields. 
The radion field is parametrized by $S$ and $T_3$.
For example, to lowest order, 
the gauge kinetic functions $f_9=S$ and $f_{5i}=T_i$, 
while the K\"ahler potential is better known \cite{ibanez}.
In the example of Ref\cite{bw}, 
with only two sets of $5$-branes (orthogonal 
to the third torus with very large $r_3=r_0$), we have 
\begin{equation}
K = -\ln(S + S^* - \sum |z^{ii}|^2) - \sum\ln(T_i + T_i^*)
 + {|z^{12}|^2 \over {2(S + S^*)^{1/2} (T_3 +T_3^*)^{1/2}}} + ...
\end{equation}
where $z^{ij}$ refers to open string chiral modes with one end of 
the string ending on the 
$i$th $5$-branes and the other end ending on the $j$th $5$-branes. (For 
$z^{ii}$ with $i=1,2$, only the $i$th torus (world-sheet) excitation modes 
are included.) 
The superpotential $W$ starts out with terms cubic in $z_i$
\begin{equation}
 W = y(S,T_i)_{jkl} z_j z_k z_l + ... 
\end{equation}
where $y(S,T_i)_{jkl}$ are model-dependent functions of the moduli.

Let us concentrate on the $F$ term $V_F$ of the effective potential.
Generically, the lowest order terms in the brane mode effective potential 
are multiplied by some functions of the moduli, while higher order 
terms couple brane modes and the moduli. 
{}From the form of the superpotential $W$, where $S$ and $T_3$
parameterize the   
radion field $r$, any brane potential will couple to the radion. 
In low orders, it will be a direct product of the brane potential and the 
radion potential.  
So the form $V_0(\psivec)[1+\fI(r)]$ is quite reasonable; this is the
first term of our assumed effective potential (\ref{eq:potent}).
Choosing $\psi$ to be the electroweak Higgs field, the last term
$V_1(\psi)$ in Eq.\ (\ref{eq:potent}) is  
simply the Higgs potential in the standard model, 
at least for $vev$ not much bigger than the electroweak scale.
Of course, we are more interested in the electroweak Higgs potential 
in the minimal supersymmetric standard model, which have two Higgs 
doublets. There, the effective Higgs potential is only poorly known. 

Notice that $V_F$ does not contain a term that involves only the
moduli, which is a property that extends to all orders in the
perturbative expansion. 
However, a term $f_0(r)$ will appear if some brane modes other than 
the inflaton develop non-zero $vev$s.
Also, we do expect effective potential terms coupling the moduli to other 
bulk modes, as well as terms 
involving the moduli to be generated non-perturbatively. Otherwise, the
moduli will appear as massless fields (much like the Brans-Dicke field), 
which is ruled out experimentally. Hence we need an effective potential 
to stabilize the radion.
This is another reason we expect the presence of
a term like $f_0(r)$. It is a bulk potential. 
This more or less justifies the choice of the form of the effective 
potential of the type (\ref{eq:potent}) proposed in the text.

\section{Brane Inflation}
\label{sec:braneinf}

The brane inflationary scenario\cite{dt} emerges rather naturally in
the generic brane world picture\cite{add,aadd,bw}. We may consider the
Type I string where $K$ branes sit more or less on top of an orientifold
plane at the lowest energy state, resulting in zero cosmological constant.
In cosmology, it is reasonable to assume that some of the
branes were relatively displaced from the orientifold plane in
the early universe. (This is the generic situation in F theory, which may 
be considered as a generalization of the Type I strings.) 
To simplify the problem, we assume that only one brane (or a set of branes)
is displaced from the rest by a distance $d$. This situation probably arises
after all except one brane have moved towards each other.
Before supersymmetry breaking and dilaton stabilization, the force 
between the separated brane and the rest is precisely zero.
In the realistic situation where supersymmetry is absent,
we expect the potential $V(d)$ to be, at large separation $d$,
\begin{equation}
 V(d) = m_s^4 d^{2-n}(1 + \sum e^{-m_i' d} - \sum e^{-m_j d})
\label{stringy}
\end{equation}
where $m_i'$ are the masses of the NS-NS string states while $m_j$ are the
masses of the string RR fields (the sums are over infinite spectra). 
For large $d$ and $n=2$, $V(d)$ is
essentially a constant. The $"1"$ term is due to gravitational interaction,
the only long range force present at large $d$.
For small $d$, the form of $V(r)$ depends crucially on the
mass spectrum.

A key feature of brane inflation is the identification of the 
separation $d$ with the vacuum expectation value of an appropriate
Higgs field\cite{witten}.
This Higgs field is an open string state with
its two ends stuck on two separated branes. That is, this Higgs field is
a brane mode playing the role of the inflaton.
In the effective four-dimensional theory, the motion of
the branes is described by this {\it slowly-rolling} scalar field,
the inflaton $\psivec = m_s^2 d$, which is the scalar component of one of the 
chiral field $z_i$, or some linear combination.
To be specific, we shall at times consider the $n=2$ case, and,
as an illustration, keep only the graviton and one RR mode, resulting in an
(over-)simplified effective potential,
\begin{equation}
 V(\psi) \sim m_s^4 (1 - e^{- |\psi _1 | /m_I}) F(\psi) + V_1(\psi _2)
\end{equation}
where $\psi_1$ and $\psi_2$ are two different brane modes and
$m_I$ is a model-dependent mass scale, which is related to the mass of
the RR mode $m_{RR}$ via $\psi _1 = m_s^2 d$, that is, $m_I m_{RR} = m_s^2$.
We also include a generic smooth function $F(\psi)$, which will be 
neglected in the text.
Since the RR mode is massless before supersymmetry breaking, and that
the supersymmetry breaking scale is below the string scale, we expect
$m_I > m_s > m_{RR}$.

In this scenario, it is even possible that the electroweak Higgs field
plays the role of the inflaton, a particularly interesting scenario.
In this case, we may identify $\psi =\psi_1 =\psi_2$ as the electroweak
Higgs field and $V_1(\psi)$ as the electroweak Higgs effective potential.
That is, $V_1(\psi = 0) \sim m_{EW}^4$.

\section{Derivation of effective low-energy description}
\label{sec:jordein}

In this appendix we derive the low energy, 4 dimensional description
given in Sec.\ \ref{sec:setup} above from the higher dimensional
description of the brane world scenario.  We start with the action
\begin{eqnarray}
L &=& 
\int d^{s+1} x^A \sqrt{-{\rm det} ( g_{AB})} \left[ {\kappa_s \over G_{(s)}} 
{}^{(s)}R  \right]  
%\nonumber \\ \mbox{} && 
+ \int d^4 y^\mu \sqrt{- {\rm det}(g_{\mu\nu})} \bigg[ {\cal L}_{\rm
b}[g_{\mu\nu}(y^\lambda), \chi(y^\lambda)] 
%\nonumber \\ \mbox{} && 
\bigg], 
\label{action0}
\end{eqnarray}
where
\beq
\kappa_s = {(s-2) \Gamma(s/2) \over 4 (s-1) \pi^{s/2} }.
\eeq
The notation here is as follows.  The number of spatial
dimensions is $s = n + 3$, $n$ is the  
number of extra compactified dimensions, and 
$G_{(s)}$ is the $s$-dimensional Newton's constant.  
The normalization
of the first term in the action (\ref{action0}) is chosen such that
the force law at short  
distances is $F = G_{(s)} m_1 m_2 / 
r^{s-1}$.  
The quantities $x^A$ are coordinates in the higher dimensional space
(the bulk) with $0 \le A \le 3+n$, $g_{AB}= g_{AB}(x^C)$ is the bulk
metric, and ${}^{(s)}R$ is the Ricci scalar of
$g_{AB}$.
In the second term, the quantities $y^\mu$ with $0 \le \mu \le 3$ are
coordinates on the brane.  The induced metric on the brane is
\beq
g_{\mu\nu}(y^\lambda) = g_{AB}[z^C(y^\lambda)] {\partial z^A \over
\partial y^\mu} {\partial z^B \over \partial y^\nu},
\eeq
where the location of the brane is $x^A = z^A(y^\mu)$.
The quantity ${\cal L}_{\rm b}$ in (1) is the Lagrangian of all the fields,
collectively called $\chi$, that live on the brane.  

This action (\ref{action0}) is a
functional of the ($4+n$) dimensional metric, of the 
location of the brane, and of the standard model fields $\psi$,
and is invariant under transformations of both
the $x^A$ coordinates and the $y^\mu$ coordinates.  
We now specialize the $x^A$ coordinate system as follows.
Lets write 
$
x^A = (x^\mu, x^a),
$
where $0 \le \mu \le 3$ and 
$4 \le a \le 3+n$.  
We can
choose the coordinate system such that the brane location is
\beq
x^A = z^A(y^\mu) = (y^\mu, 0, 0, \ldots, 0).
\label{(braneloc)}
\eeq
Hence we can identify the first four of the bulk coordinates $x^\mu$
with the brane coordinates $y^\mu$.

We now make the ansatz for the metric, in the above choice of bulk
coordinates $x^A$, of
\begin{equation}
\label{metric-ansatz}
g_{AB}(x^\mu,x^a)  = \left[ \begin{array}{cc} 
g_{\alpha\beta}(x^\mu)   & 0 \\
0 & e^{2 {\tilde \Phi}(x^\mu)} h_{ab}(x^c) \end{array} \right].
\end{equation}
Here it is assumed that the internal space is a compact space of constant
curvature, like an $n$-sphere $S^n$ or an $n$ torus $S^1 \times \ldots
\times S^1$, with metric $h_{ab}(x^c)$.  The volume $V_n$ and
effective radius $r$ of 
the extra dimensions are then given by
\beq
V_n = r^n = e^{n {\tilde \Phi}} \, \int d^n x^a \, \sqrt{{\rm
det}(h_{ab})}.
\eeq
If $r_0$ is the value of the radius $r$ today, lets adopt the
convention that
\beq
\int d^n x^a \, \sqrt{{\rm det}(h_{ab})} = r_0^n,
\label{h-convention}
\eeq
so that 
\beq
r = e^{{\tilde \Phi}} r_0.
\label{rPhi0}
\eeq

In going from the full metric $g_{AB}(x^\mu,x^a)$ to the reduced form
(\ref{metric-ansatz}) we have thrown away all the Kaluza-Klein modes
which have masses $\agt 1/r$.  Hence the ansatz (\ref{metric-ansatz})
will only be valid when all the fields vary with $x^\mu$ over length
scales $\gg r$.  We have also thrown away several of the components of
the metric -- the components $g_{\mu\,a}(x^\mu)$ and the traceless
part of $g_{ab}(x^\mu)$.  This is valid since these components have no
couplings to the brane fields $\psi$; in the four dimensional
description they will act as free, massless
scalar and vector fields which are coupled only to the metric
$g_{\mu\nu}$ \footnote{To see that these fields are exactly decoupled
classically, one should 
use a definition of the radion field ${\tilde \Phi}$ which is more
general than Eq.\ (\ref{metric-ansatz}), namely
$$
\int d^n x^a \, \sqrt{ - {\rm det} (g_{AB}) } = r_0^n e^{n {\tilde
\Phi}} \sqrt{ - {\rm det} 
(g_{\mu\nu})}.  
$$
This equation together with the action (\ref{action0}) shows that
${\tilde \Phi}$ is the only piece of the metric which has couplings to
anything other than the metric.}
Their equations of motion will be source
free equations of the form $\nabla_\alpha \nabla^\alpha \varphi =0$
\cite{Golda}, and so, at least classically, we can take them to vanish 
\footnote{Quantum mechanically, these fields will be subject to
the same process of parametric amplification during inflation as
normal gravitons, and if they start 
in their vacuum states the total $\Omega$ in these fields today should
presumably be comparable to the total $\Omega$ in relic gravitons from
inflation, which is of the order of $10^{-14}$ in typical inflation
models but smaller in the models of this paper.}.

We now specialize the brane Lagrangian ${\cal L}_b$ appearing in Eq.\
(\ref{action0}) to be of the form
\beq
{\cal L}_b(g_{\mu\nu},\chi) = - {1 \over 2} ( \nabla \psivec)^2 + 
{\cal L}_{\rm rest}(g_{\mu\nu},\chi_{\rm rest}),
\label{eq:braneaction}
\eeq
where $\psivec$ is the inflaton field or fields, and 
$\chi_{\rm rest}$ denotes the remaining brane fields other than
$\psi$, described by the Lagrangian ${\cal L}_{\rm rest}$.
We also add to the action the terms
\beq
- \int d^{s+1} x^A \sqrt{-{\rm det} ( g_{AB})} \ V_{\rm bulk}({\tilde
 \Phi}) - \int d^{4} x^a \sqrt{-{\rm det} ( g_{\mu\nu})} \ V_{\rm
brane}(\psivec,{\tilde \Phi}).
\label{bulk-pot}
\eeq
This consists of a bulk potential energy per unit s-dimensional volume
$V_{\rm bulk}$ and a brane potential energy per unit 3-volume $V_{\rm
brane}$ \footnote{Note that the explicit dependence of these
potentials on the metric component ${\tilde \Phi}$ spoils the
covariance of the full action under transformation of the $x^A$
coordinates; it is difficult to write down a fully covariant
radius-stabilization mechanism.}.  We discussed in Appendix
\ref{sec:effpot} above the physical origin for such terms which depend
on the size of the extra dimensions as well as on the inflaton.

Using the ansatz (\ref{metric-ansatz}) in the action (\ref{action0}),
inserting the brane action (\ref{eq:braneaction}) 
and adding the potential terms (\ref{bulk-pot}) now yields the reduced
4-dimensional action
\begin{eqnarray}
S &=& \int d^4 x \sqrt{\det g} \, \bigg[ { e^{n {\tilde \Phi}} 
\over 16 \pi G} {}^{(4)}R + { n (n-1) \over 16 \pi G } e^{n {\tilde
\Phi}} (\nabla {\tilde \Phi})^2 
%\nonumber \\ \mbox{} && 
- V(\psivec,{\tilde \Phi}) + 
{\cal L}_{\rm sm}(g_{\mu\nu}, \psi) \bigg].
\label{action_Jordan}
\end{eqnarray}
Here $G$ is the usual 3-dimensional
Newton's constant, given by
\beq
{1 \over 16 \pi G} = 
{\kappa_s r_0^n \over  G_{(s)}}, 
\eeq
where $r_0$ is the equilibrium value of the radius of the extra
dimensions.
The action (\ref{action_Jordan}) has the form of a scalar-tensor
theory of gravity, written in the 
Jordan frame.  Note that the sign of the kinetic term
for the scalar field in the action (\ref{action_Jordan}) is opposite
to the normal sign; this is not a problem since it is the sign of the
kinetic energy term in the Einstein frame (see below) that is relevant
to considerations like stability and positivity of energy etc.
The Jordan-frame potential $V$ is given by
\begin{eqnarray}
V(\psivec,{\tilde \Phi}) &=& r_0^n e^{n {\tilde \Phi}} V_{\rm bulk}({\tilde
\Phi}) - { k_i \over 8 \pi G r_0^2 } \, e^{(n-2) {\tilde \Phi}}
%\nonumber \\ \mbox{} && 
+ V_{\rm brane}(\psivec,{\tilde \Phi}),
\label{potentialJ}
\end{eqnarray}
where the Ricci scalar of the metric $h_{ab}$ is $2 k_i r_0^{-2}$ and
$k_i$ is a dimensionless constant of order unity (cf Eq.\
(\ref{h-convention}) above).  From now on we specialize to flat
internal spaces so that $k_i=0$.  Then we see that only the particular
combination $V = r^n V_{\rm bulk} + V_{\rm brane}$ of the potentials
$V_{\rm bulk}$ and $V_{\rm brane}$ is relevant in the low energy
description.  Our assumed form for this potential $V$ is given in Eq.\
(\ref{eq:potent}) above.

Finally, we transform to the Einstein frame description.  
We introduce a canonically normalized radion field by defining $\mu$
according to Eq.\ (\ref{mudef}) above, and we define
\beq
\Phi = n \mu {\tilde \Phi}.
\eeq
The radius $r$ is thus related to $\Phi$ by 
\beq
r = r_0 \exp \left[{\Phi \over  n \mu }\right]
\label{rPhirelation}
\eeq
from Eq.\ (\ref{rPhi0}).  We define the Einstein frame metric by
\beq
{\hat g}_{\alpha\beta} = e^{\Phi / \mu} g_{\alpha\beta} = e^{n {\tilde
\Phi}} g_{\alpha\beta}.
\label{Emetric}
\eeq
The action then takes the form of Eq.\ (\ref{action2}) above,
where
\beq
S_{\rm rest}[g_{\alpha\beta},\chi_{\rm rest}] \equiv \int d^4 x \sqrt{- { g}}
{\cal L}_{\rm rest}(g_{\alpha\beta}, \chi_{\rm rest})
\eeq
is the action of the brane fields $\chi_{\rm rest}$.

The equations of motion derived from the action (\ref{action2}), when
we treat the last term as a fluid with Jordan-frame density $\rho$ and
pressure $p$, are
\begin{eqnarray}
{\mpl^2 \over 8 \pi} {\hat G}_{\alpha\beta} &=& 
{\hat \nabla}_\alpha \Phi {\hat \nabla}_\beta \Phi - {1 \over 2}
{\hat g}_{\alpha\beta} ({\hat \nabla} \Phi)^2 + 
e^{-\Phi/\mu} \left[
{\hat \nabla}_\alpha \psi {\hat \nabla}_\beta \psi - {1 \over 2}
{\hat g}_{\alpha\beta} ({\hat \nabla} \psi)^2 \right] \nonumber \\
\mbox{} && - e^{-2 \Phi / \mu} V(\psi,\Phi) {\hat g}_{\alpha\beta} +
e^{-2 \Phi / \mu} \left[ ({\rho} + {p}) {\hat u}_\alpha
{\hat u}_\beta + p {\hat g}_{\alpha\beta} \right],
\end{eqnarray}
\beq
{\hat \nabla}_\alpha {\hat \nabla}^\alpha \Phi + {1 \over 2 \mu}
e^{-\Phi / \mu} ({\hat \nabla} \psi)^2 - {\partial \over \partial
\Phi} \left[ e^{-2 \Phi / \mu} V(\psi,\Phi) \right] + {1 \over 2 \mu}
e^{- 2 \Phi / \mu} (\rho - 3 p) =0,
\eeq
and
\beq
{\hat \nabla}_\alpha {\hat \nabla}^\alpha \psi - {1 \over  \mu}
{\hat \nabla}_\alpha \Phi \, {\hat \nabla}^\alpha \psi 
- e^{-\Phi / \mu} {\partial  \over \partial \psi} V(\psi,\Phi)=0.
\eeq
Here ${\hat \nabla}_\alpha$ is the derivative operator associated with
the Einstein frame metric ${\hat g}_{\alpha\beta}$, and ${\hat
u}_\alpha$ is normalized with respect to ${\hat g}_{\alpha\beta}$.

\section{A particular realization of Phase 0 Evolution}
\label{sec:phase0more}

Here, we study the pre-inflation Phase 0 in some detail
for the sinusoidal potential 
\beq
f_I(r)=a(1-\cos mr)
\label{fdef}
\eeq
introduced in \S\ref{sec:phase0}.
For this choice, the effective potential for the radion field
$\Phi$ in the Einstein
frame is, from Eq.\ (\ref{F0}), 
\be
\veff=\Vhatz\biggl({\rzero\over r}\biggr)^{2n}
(1+a-a\cos mr).
\ee
If $amr/2n\gg 1$, it is easy to show that the maxima of $\veff$ are
at
\be
mr_j=(2j+1)\pi+\epsilon_j\qquad\qquad
\epsilon_j\approx {2n(1+2a)\over amr_j}.
\ee
The heights of successive maxima of $\veff$ differ by approximately
\be
\veff^{(j+1)}-\veff^{(j)}\approx
-2n\Vhatz(1+2a)\biggl({\rzero\over r_j}\biggr)^{2n}
\biggl({r_{j+1}-r_j\over r_j}\biggr);
\ee
the potential is biased toward large values of $r$.
Since whether or not the radion escapes over the first barrier is determined
at very early times, we can ignore expansion, and treat the dynamics as
completely conservative. If the radion starts out with zero or negligible
kinetic energy, then if it can only escape over the barrier at $r_{j+1}$
if it begins sufficiently close to $r_j$. Expanding the effective
potential near $r_j$ we find that
\be
\veff(r_j+\Delta r_j)-\veff^{(j)}\approx -2na\Vhatz\biggl({\rzero\over r_j}\biggr)^{2n}
(m\Delta r_j)^2,
\ee
$\veff(r_j+\Delta r_j)=\veff(r_{j+1})$
when 
\be
m\Delta r_j\approx\sqrt{8\pi n(1+2a)\over amr_j},
\ee
which is small if $amr_j/2n\gg 1$. If the radion begins at any radii
$r_j+\Delta r_j\lta r\lta r_{j+1}$ it should be captured at the nearest
minimum, which is roughly halfway between $r_j$ and $r_{j+1}$; if it
begins at $r_j\leq r\lta r_j+\Delta r_j$, then it should grow without
bound. Numerical solutions of eqs. (\ref{eq:phase01}) verify this
picture. If the value of $r$ at the beginning of Phase 0 is random,
the probability that the radion is {\it not} trapped is 
$P_e\approx m\Delta r_j/2\pi$, which {\it decreases}
$\propto (2n/amr)^{1/2}$ as $amr/2n$ grows. Thus, for large $amr/2n$,
it is extremely likely, although {\it not guaranteed},  that the
radion is trapped at the minimum of $\veff$ nearest to its starting
value of $r$.

The Einstein-frame expansion rate once the radion settles into a
minimum of $\veff(\Phi)$ is 
\be
\Hbar^2\approx{8\pi\Vhatz\over 3\mpl^2}\exp(-2\Phi/\mu),
\ee
where $\Phi$ is the value of the field. Expanding $\veff(\Phi)$
around the minimum implies
\be
\veff(\Phi+\Delta\Phi)\approx a\Vhatz\biggl({mr\over n\mu}\biggr)^2
\exp(-2\Phi/\mu){(\Delta\Phi)^2\over 2},
\ee 
so the characteristic oscillation frequency for fluctuations in
the radion field is
\be
\omega_\Phi^2\approx a\biggl({mr\over n\mu}\biggr)^2 
\exp(-2\Phi/\mu)\approx{12a(mr)^2\Hbar^2\over n(n+2)}.
\label{eq:iralm}
\ee
For large values of $mr$ (and $a$ not too small), $\omega_\Phi\gg
\Hbar$. One consequence of this inequality is that we do not
expect large-scale fluctuations in the radion field to arise
during inflation.

\end{document}